\documentclass[12pt]{article}

\IfFileExists{srcltx.sty}{\usepackage[active]{srcltx}}

\usepackage{amssymb,amsmath,array,varioref}%
\usepackage{graphicx,subfig}%

\newcommand{\eV}{\ensuremath{\:\mathrm{eV}}} %
\newcommand{\kev}{\ensuremath{\:\mathrm{keV}}} %
\newcommand{\mev}{\ensuremath{\:\mathrm{MeV}}} %
\newcommand{\gev}{\ensuremath{\:\mathrm{GeV}}} %
\newcommand{\keV}{\ensuremath{\:\mathrm{keV}}} %
\newcommand{\MeV}{\ensuremath{\:\mathrm{MeV}}} %
\newcommand{\Neff}{\ensuremath{N_\text{eff}}\xspace}
\newcommand{\MassThresh}{\ensuremath{40\:\MeV}}

 \usepackage{xspace,paralist}
\usepackage[a4paper,hscale=.72, vscale=.72]{geometry}

\usepackage{url}

\newcommand{\numsm}{$\nu$MSM\xspace}
\newcommand{\p}{\partial}

\newcommand{\diag}{{\rm diag}}

\begin{document}

\title{\rightline{\small \tt CERN-PH-TH-2012-042} Restrictions on the lifetime
  of sterile neutrinos from primordial nucleosynthesis}

\author{Oleg Ruchayskiy\thanks{CERN Physics Department, Theory Division,
    CH-1211 Geneva 23, Switzerland}~ and Artem
  Ivashko\thanks{Instituut-Lorentz for Theoretical Physics, Universiteit
    Leiden, Niels Bohrweg 2, Leiden, The Netherlands}
\thanks{Department of Physics, Kiev National Taras Shevchenko University, Glushkov str. 2 building 6, Kiev, 03022,
Ukraine}}

\date{}

\maketitle

\begin{abstract}
  We analyze the influence of sterile neutrinos with the masses in the MeV
  range on the primordial abundances of Helium-4 and Deuterium. We solve
  explicitly the Boltzmann equations for all particle species, taking into
  account neutrino flavour oscillations and demonstrate that the abundances
  are sensitive \emph{mostly} to the sterile neutrino lifetime and only weakly
  to the way the active-sterile mixing is distributed between flavours. The
  decay of these particles also perturbs the spectra of (decoupled) neutrinos
  and heats photons, changing the ratio of neutrino to photon energy density,
  that can be interpreted as extra neutrino species at the recombination
  epoch. We derive upper bounds on the lifetime of sterile neutrinos based on
  both astrophysical and cosmological measurements of Helium-4 and Deuterium.
  We also demonstrate that the recent results of Izotov \&
  Thuan~\cite{Izotov:2010ca}, who find $2\sigma$ higher than predicted by the
  standard primordial nucleosynthesis value of Helium-4 abundance, are
  consistent with the presence in the plasma of sterile neutrinos with the
  lifetime 0.01 -- 2 seconds.
\end{abstract}

\section{Introduction}
\label{sec:introduction}
The characteristic feature of the physical processes in the early Universe is
a peculiar interplay of gravity and microscopic physics. Gravity introduces
the \emph{Hubble time parameter} $\tau_H$ that indicates the timescale on
which the global properties of the Universe (geometry, temperature, etc.)
change significantly. The Hubble time is determined solely by the energy
density of the matter filling the space. The microscopic matter constituents,
particles, are involved in the interaction processes, that are believed to be
described fundamentally by three known forces --- electromagnetic, weak and
strong. As long as the timescale $\tau$ of any given microscopic physical
process is much smaller than $\tau_H$, the expansion can be neglected on that
timescale. If time $\tau$ is enough to establish thermal equilibrium between
the particles, then the equilibrium it maintained in the course of the
Universe expansion, while $\tau \ll \tau_H$ holds. When this inequality ceases
to hold, the state of equilibrium is lost. The main reason for that is that
interparticle distances become larger, while the corresponding densities
become lower, hence interactions are less likely to occur.

In this paper we are consider the formation of light nuclei in the primordial
environment -- Big Bang nucleosynthesis (BBN). All three fundamental
interactions are important for this phenomenon, all playing different
roles. Charged particles together with photons are subject to electromagnetic
forces and the equilibration timescale of corresponding processes is tiny with
respect to the expansion time. Therefore the particles are kept in thermal
equilibrium at the common temperature $T$. Due to expansion the temperature is
\textit{decreasing} with time. The equilibration time of the weak interactions
changes abruptly so that at $T \gtrsim\text{few}\MeV$ weakly interacting
neutral particles (neutrinos and neutrons) stay in equilibrium, while at lower
temperatures they fall out of it (freeze out).

At high temperatures processes like $n+\nu_e\rightarrow p + e^-$ maintain
\emph{chemical} equilibrium, that is the neutron-to-proton conversion exhibits
the same finite intensity as the opposite processes.  Chemical and thermal
equilibria are interconnected, so they are lost simultaneously, when
neutron-to-proton ratio freezes out.  Finally, the strong interactions are
responsible for the production of nuclei comprising more than one nucleon. The
most important fusion reaction for the formation of the first nucleus,
deuteron, $n+p \rightarrow D$, \emph{releases} energy of at least the binding
energy of deuteron $E_D \approx 2.2 \MeV$, and proceeds effectively in dense
primordial medium. At temperatures of the order of $E_D$, however, energetic
photons collide with deuteron and lead to its destruction. As baryon density
is much lower than the density of photons~\cite{WMAP7}, there are \emph{many}
photons with energies much higher than $E_D$ that collide with deuterons and
hence postpone the production of the significant deuteron density until the
temperature when the photodissociation is not effective anymore, $T\simeq
80\keV$, much lower than the binding energy. The net abundance of deuterium
is, however, non-zero at all times till this moment and is given by the
equilibrium Boltzmann distribution. Deuterium that is created at lower
temperatures, serves as a fuel for the formation of $^3$He, $^4$He and other
nuclides.

Although the times of elements' production and the moment of the departure
from the chemical $p-n$ equilibrium are well-separated, the former process is
very sensitive to the latter. Firstly, the details of the freeze-out set the
ratio of the neutron to proton densities, and secondly, the time elapsed
between the two moments determines the fraction of neutrons that have decayed
since then (recalling that neutron is an unstable particle).

The seminal ideas of the primordial synthesis of light elements were first
outlined in the so-called $\alpha\beta\gamma$ paper, \cite{Alpher:1948ve},
published in the late 1940s.  Since then the theory of Big Bang
nucleosynthesis has evolved and its main predictions were confirmed, making it
a well-developed model from both theoretical and observational points of
view. A lot of reviews of the standard BBN scenario and its implication for
particle physics models exist (see
e.g.~\cite{Iocco:08,Steigman:07,Pospelov:10b}).

The predictions of the primordial nucleosynthesis can change once one replaces the
Standard Model of particle physics underlying the processes considered so far
by some of its ``beyond the Standard Model'' (BSM) extensions. Therefore the
BBN plays the role of a benchmark for testing physical models. 

\begin{figure}[!tp]
  \centering \subfloat[Quadratic mixing $\nu_S\leftrightarrow \nu_e$ of sterile
  neutrino with
  $\nu_e$]{\includegraphics[width=.3\textwidth]{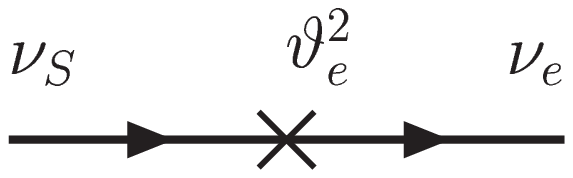}}\newline
  \subfloat[Decay of sterile neutrino $\nu_S \to \nu_e \nu_\alpha \bar\nu_\alpha$
  through neutral current
  interactions]{\includegraphics[width=.4\textwidth]{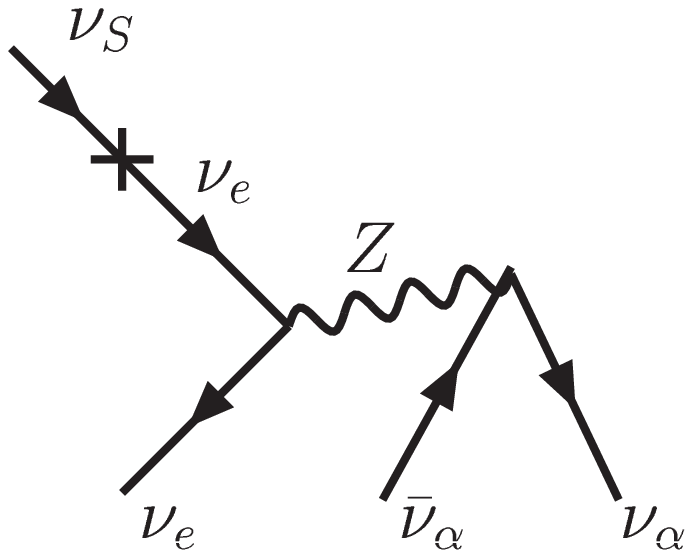}}
  ~~~~~~~~~~~~ \subfloat[Fermi-like interaction with the ``effective'' Fermi
  constant $\vartheta_e\times G_F$ for the process in the panel (b).]
  {\includegraphics[width=.4\textwidth]{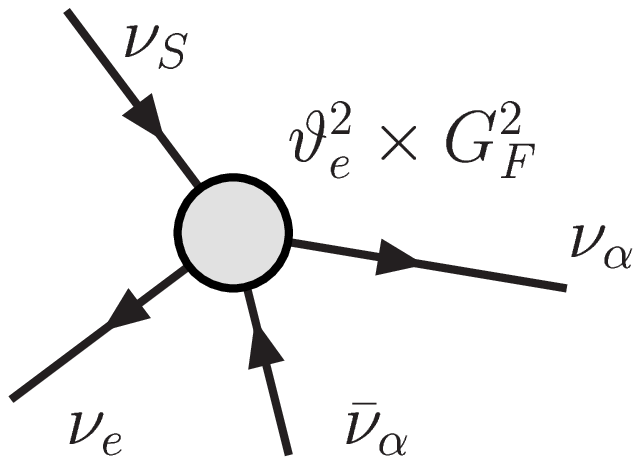} }
  \caption{Fermi-like super-weak interactions of sterile neutrino}
  \label{fig:fermi1}
\end{figure}

In this paper we investigate the influence of \emph{sterile neutrinos} on
primordial nucleosynthesis. Sterile neutrinos are hypothetical massive
\textit{super-weakly-interacting} particles (see
e.g.~\cite{Boyarsky:09a,Kusenko:09a} for reviews), as opposed to their
weakly-interacting counterparts -- ordinary Standard Model neutrinos
$\nu_e,\nu_\mu, \nu_\tau$, that are called ``\emph{active}'' in this context.
Sterile neutrinos carry no charges with respect to the Standard Model gauge
groups (hence the name), but via their quadratic \emph{mixing} to active
neutrinos they effectively participate in weak reactions and at energies much
below the mass of the $W$-boson their interaction can be described by the
analog of the Fermi theory with the Fermi coupling constant $G_F$ replaced by
$G_F\times \vartheta_\alpha$, where the \emph{active-sterile mixing angle}
$\vartheta_\alpha \ll 1$ (see Fig.~\ref{fig:fermi1}).  Here $\alpha$ is a
flavour index, $\alpha = e,\mu,\tau$, indicating that sterile neutrino can mix
differently with neutrinos of different flavours. Massive sterile neutrinos
can decay, but due to their feeble interaction strength their lifetime can be
of order seconds (even for masses as large as MeV). The decay products of the
sterile neutrinos are injected into the primordial environment, increasing its
temperature and shifting the chemical equilibrium. 

In this work we concentrate on sterile neutrinos with the masses in the MeV
range, motivated by the recent
observations~\cite{Akhmedov:98,Asaka:05a,Asaka:05b,Shaposhnikov:08a,Canetti:10a}
that particles with such masses can be responsible simultaneously for neutrino
oscillations and generation of baryon and lepton asymmetry of the Universe and
can influence the subsequent generation of dark matter~\cite{Laine:08a}. The
corresponding model has been dubbed $\nu$MSM (\emph{Neutrino Minimal Standard
  Model}, see~\cite{Boyarsky:09a} for review).

Several works had previously considered the influence of MeV-scale particles
on primordial nucleosynthesis. Compared to the
Refs.~\cite{Dolgov:00a,Dolgov:00b} this paper accounts for the neutrino
flavour oscillations in the plasma and employs more accurate strategy of
solving Boltzmann equations, which results in the revision of the bounds
of~\cite{Dolgov:00a,Dolgov:00b} (see Section~\ref{sec:discussion} for detailed
comparison). The authors of~\cite{Smith:08} developed a new code that can
perform treatment of active and sterile neutrinos with arbitrary distribution
functions, non-zero lepton asymmetry, etc.  However, as of time of writing
this code has not been made publicly available and the Ref.~\cite{Smith:08}
did not derive bounds on sterile neutrino parameters. The
work~\cite{Fuller:2011qy} concentrated on the bounds that cosmic microwave
background measurements could provide on decaying sterile neutrinos with the
masses $100-500$~MeV, leaving BBN analysis for the future work. A number of
other works
(\cite{Dolgov:1997it,Dolgov:88,Kawasaki:1993gz,Kawasaki:2000en,Hannestad:04})
analyzed the influence of decaying MeV particles on BBN. We compare with them
in the corresponding parts of the paper.

\paragraph{The paper is organized as follows.} We explain the modifications of
the standard BBN computations due to the presence of sterile neutrinos in the
plasma and describe our numerical procedure in Sec.~\ref{sec:bbn-formalism}.
The results are summarized in Sec.~\ref{sec:results}. We conclude in
Sec.~\ref{sec:discussion}.
Appendixes~\ref{sec:numerical-tests}--\ref{app:neutrino-oscillations} provide
the details of our numerical procedure.

\section{Big Bang Nucleosynthesis with sterile neutrinos}
\label{sec:bbn-formalism}

The section below summarizes our setup for the BBN analysis with decaying
particles. The notations and conventions closely follow the series of
works~\cite{Dolgov:00b,Dolgov:98addendum,Dolgov:97sm}.

We will be interested only in the tree-level Fermi interactions of sterile
neutrinos with the primordial plasma. In this case the interaction is fully
determined by the squares of their mixing angles.  We will consider one
Majorana particle with 4 degrees of freedom\footnote{This number corresponds
  to $g_s = 2$ of additional chiral singlets (i.e. ``neutrino-like''
  species). Actual number of degrees of freedom is of course twice larger:
  $\text{dof}=2\times g_s =4$, because every chiral fermion has 2 different
  helicity states.} and three \emph{active-sterile mixing angles}
$\vartheta_\alpha^2 $. Matrix elements of interactions of sterile neutrinos
with the Standard Model particles are summarized in Appendix~\ref{app:matrix-elements}
(Tables~\ref{tab:sterile-scatterings} -- \vref{tab:sterile-decays}).

We consider in this work only sterile neutrinos with the masses in the range
$1\MeV<M_s<M_\pi\approx 140\MeV$. For heavier particles, two-particle decay
channels appear (e.g. $\nu_S \to \pi_0 \nu_\alpha,\pi^\pm e^\mp$) and our procedure
of solving Boltzmann equations (described below) should be significantly
modified. The lower bound was chosen to be around $1\mev$ by the following
considerations. The
sterile neutrino lifetime $\tau_s$ is \cite{Gorbunov:07a}
\begin{equation}
\label{eq:lifetime-expression}
\begin{aligned}
  \tau_s^{-1} = \Gamma_s & = \frac{G_F^2 M_s^5}{96\pi^3}\left[(1+\tilde g_L^2+
    g_R^2)(\vartheta^2_\mu+\vartheta^2_\tau)+(1+g_L^2+g_R^2)\vartheta^2_e \right]  \\
  &\approx 6.9~\mathrm{sec}^{-1}~\left(\frac{M_s}{10\mev}\right)^5\Bigr[
  1.6\,\vartheta_e^2 + 1.13(\vartheta_\mu^2 + \vartheta_\tau^2)\Bigr]
\end{aligned}
\end{equation}
where $\theta_W$ is the Weinberg's angle and $g_R=\sin^2 \theta_W \approx
0.23$ , $g_L=\frac12+\sin^2 \theta_W$, $\tilde{g}_L=-\frac12+\sin^2
\theta_W$.\footnote{The expression~(\ref{eq:lifetime-expression}) is for
  Majorana particle. For Dirac particle the lifetime would be twice larger.}
From this expression one sees that sterile neutrinos lighter than about
$2\MeV$ have lifetime of at least several hundred seconds even for very large
mixing angles $\vartheta\sim 1$.
Therefore, such particles survive till the onset of the BBN, and freeze-out at
temperatures $T\sim 2-3\MeV$.  They would be relativistic at that time, 
  i.e.\ their average momentum would be of the order of temperature, 
  $\langle p \rangle \sim T$, and their contribution to the number of
  relativistic neutrino species would be significant, $\Delta\Neff \simeq
  2$. In the course of the Universe expansion $\langle p \rangle $ would scale
  as temperature due to the gravitational redshift, and at some point would
  become smaller than the mass of sterile neutrino. At that moment the energy
  density of sterile neutrinos would start to change with expansion as
  $a^{-3}$ rather than $a^{-4}$ (where $a$ is a scale-factor) so that the
  contribution of these massive particles to the energy density would quickly
  become dominant, making $\Neff \gg 1$ (or could even overclose the Universe)
  before the production of light elements starts. It contradicts the current
  bound that puts $\Neff = 3.74^{+0.8}_{-0.7}\pm 0.06(\text{syst})$ at
  $2\sigma$~\cite{Izotov:2010ca}.\footnote{Here the systematic error is
    due to the different values of neutron lifetime between the average value
    from Particle Data group,~\cite{Nakamura:2010zzi} and the recent
    measurement of~\cite{Serebrov:2010sg}.}

Additionally, in the $\nu$MSM the successful baryogenesis is possible only for
the masses of sterile neutrinos above few MeV~\cite{Asaka:05b,Canetti:10a}.
Therefore we restrict the analysis to the region of masses higher than
$1\MeV$.

\subsection{Expanding Universe and distributions of particles}
\label{sec:expansion-dynamics}
We consider expansion of the homogeneous and isotropic Universe with the flat
Friedmann--Robertson--Walker metric in the form $ds^2 = dt^2 - a^2
d\vec{x}^2$, where $a=a(t)$ is a time-dependent scale factor, whose evolution
is described by the Friedmann equation
\begin{equation}
\label{eq:Friedmann-first}
H \equiv \frac{\dot a}{a}=\sqrt{ \frac{8\pi G_N}{3} \rho~}~,
\end{equation}
with the quantity on the left-hand side being the Hubble expansion rate,
reciprocal to the expansion timescale $\tau_H$ discussed above.  The total
energy density $\rho$ is the sum of all the energy densities present in the
medium, and $G_N$ is the Newton's constant. The energy density together with
the total pressure density $p$ satisfy the ``energy conservation'' law
 \begin{equation}
   \label{eq:Friedmann-second}
   a\frac{d\rho}{da} + 3 (p+\rho)=0.
 \end{equation}
 At the temperatures of interest the dominant components of the plasma are
 photons $\gamma$, electrons and positrons $e^\pm$, three flavours of active
 neutrinos ($\nu_e, \nu_\mu, \nu_\tau$) and sterile neutrinos.\footnote{Muons
   may appear in plasma from the decays of the sterile neutrinos with $M_s >
   106\MeV$. See Sec.~\ref{sec:sterile-dynamics} for details.} Working with
 the particle kinematics in the expanding Universe it is convenient to use
 \emph{conformal momentum $y$} instead of the usual physical momentum $p$. The
 two are related through $y=pa$.
 The quantitative description of the plasma population is provided by the
 distribution functions $f_\alpha$, that are the numbers of particles $\alpha$
 per ``unit cell'' of the phase space $d^3 p\; d^3 x = (2\pi)^3$.  At keV--MeV
 temperatures the medium is homogeneous and the distribution functions are
 independent of spatial coordinates of particles, and due to isotropy they do
 not depend on the direction of the particle momentum. That simplifies the
 description of their evolution and therefore
 \begin{equation}
   \label{eq:5}
   \frac{df}{dt} \equiv \left(\frac{\p f}{\p t} - Hp \frac{\p f}{\p p}\right)
   = \frac{\p f (t,y)} {\p t}
 \end{equation}
 holds.  The goal is to find the distribution functions of all relevant
 particles and to use them to compute the energy density and pressure as a
 function of time and scale-factor, closing the system of
 Eqs.~(\ref{eq:Friedmann-first})--(\ref{eq:Friedmann-second}) via
\begin{equation}
  \label{eq:1}
  \rho=\sum_i \frac{g_i}{2\pi^2} \int f_i {E}_i p^2
  dp\quad;\quad
  p=\sum_i \frac{g_i}{6\pi^2} \int f_i \frac{p^4}{{E}_i} dp\
\end{equation}
Here the summation goes over all plasma particles, $g_i, m_i$ is the number of
degrees of freedom and mass of $i$-th particle respectively, ${E}_i = \sqrt{p^2 +m_i^2}$.

 If interaction rate of the particles is much faster than the Hubble expansion
 rate, their distribution functions are given by either the Bose-Einstein, or
 the Fermi-Dirac distributions. This is the case for photons, electrons and
 positrons --- that are kept in equilibrium due to intensive electromagnetic
 interactions
 \begin{equation}
   \label{eq:equilibrium-distribution}
   f_\gamma = \frac{1}{e^{E/T}-1},~~f_e = \frac{1}{e^{E/T}+1}.
 \end{equation}
 Their contribution to the energy and pressure in
 Eqs. (\ref{eq:Friedmann-first}), (\ref{eq:Friedmann-second}) is hence
 determined by the single parameter -- temperature. However, to describe the
 contributions of the other particles one has to solve \textit{kinetic}
 equations involving them (see
 Secs.~\ref{sec:active-dynamics}--\ref{sec:sterile-dynamics} below).

\subsection{Baryonic matter}
\label{sec:baryonic-matter}
The contribution of the baryonic matter to the evolution of the hot plasma of
relativistic species is proportional to the so-called baryon-to-photon ratio
$\eta_B = n_B/n_\gamma$. The measurements of relic radiation \cite{WMAP7}
yield $\eta_B = (6.19\pm 0.15)\times 10^{-10}$. One can see that baryons are
present in negligible amount, and do not influence the dynamics of the
remaining medium.  This allows to analyze our problem in two steps. At
\textsc{step i} we omit baryonic species and study how the temperature of the
plasma, the expansion factor and neutrino distributions evolve in time from
temperatures of the order of $100\MeV$, when sterile neutrinos typically start
to go out of equilibrium,\footnote{The exact ``freeze-out'' temperature
  depends on the mixing angle.} down to $T_{\rm Fin} \simeq 10\kev$ when
nuclear fusion reactions have ended.  At \textsc{step ii} we use these results
to determine the outcome of the nuclear reaction network against the
background of evolving electromagnetic plasma
(Sec.~\ref{sec:nuclear-dynamics}).

 \subsection{Active neutrinos at MeV temperatures}
 \label{sec:active-dynamics}

 Weak interactions are not able to maintain the thermal equilibrium of active
 neutrinos with the plasma during all the expansion period we consider. A
 simple comparison of the weak collision rate $G_F^2 T^5 $ and $H(T)$ tells
 that neutrino maintain their equilibrium with the rest of the plasma down to
 temperatures $T_{\rm dec}\sim \text{few}\mev$.  The process of neutrinos
 going out of equilibrium is usually referred to as \emph{neutrino
   decoupling}. Throughout the paper we assume that \textit{no} large lepton
 asymmetry is present so that the number of neutrinos is equal to the number
 of antineutrinos.\footnote{For the previous studies of the BBN outcomes
   \textit{with} the large lepton asymmetry present see
   e.g.~\cite{Lesgourgues:99,Serpico:05,Smith:2006uw,Smith:08,Mangano:10}.} At
 temperatures higher than $T_{\rm dec}$ the distribution is therefore given by
 the Fermi-Dirac one, while at lower temperatures we have to solve the set of
 three Boltzmann equations
\begin{equation}
\label{eq:Boltzmann-equation}
\frac{df_{\nu_\alpha}}{dt}  =  I_\alpha,~~\alpha=e,\mu,\tau
\end{equation}
The details of the interactions, such as particle collisions, are encoded in
the so-called collision terms $I_\alpha$. The terms are
explicitly~\cite{Kolb:90}
\begin{equation}
  \label{eq:4}
  I_\alpha =  \frac{1}{2 E_\alpha} \sum\limits_{\rm in, out} \int  S |\mathcal{M}|^2 F[f] (2\pi)^4 \delta^4 (p_{\rm in}-p_{\rm out}) \prod\limits_{i=2}^Q \frac{d^3 p_i}{(2\pi)^3 2 E_i}
\end{equation}
The sum runs over all the possible initial states ``in'' involving
$\nu_\alpha$ (represented by a particle set $\nu_\alpha,~2,~3,~\hdots,~K$) and
the final states ``out'' ($K+1,~\hdots,~Q$). Matrix element $\mathcal{M}$
corresponds to the probability of the transition ``in''--``out'' to occur and
the delta-function ensures the conservation of 4-momentum $p_{\rm in}=p_{\rm
  out}$. Symmetrization factor $S$ is equal to $1$, except of the transitions
involving identical particles either in initial or in a final state. Relevant
matrix elements together with the symmetrization factors are listed in
Appendix \ref{app:matrix-elements}. The interaction rates are dependent on the
population of the medium, and the functional $F[f]$ describes this. In case
when all the incoming and outgoing particles are fermions,
\begin{equation}
  \label{eq:2}
  F[f]=(1-f_{\nu_\alpha})\hdots (1-f_K) f_{K+1} \hdots f_Q - f_{\nu_\alpha} \hdots f_K (1-f_{K+1}) \hdots (1-f_Q).
\end{equation}   
When some of particles are bosons, one has to replace $(1-f_R)$ by $(1+f_R)$ for every bosonic particle $R$. 

A simple estimate (see Appendix~\ref{app:neutrino-oscillations}) demonstrates
that the rates of transitions between neutrinos of different flavours are much
faster than weak reactions.  We argue 
that this phenomenon can be approximately described by the following
modification of the Boltzmann equations
\begin{equation}
\label{eq:Boltzmann-equation-oscillations}
\frac{d f_{\nu_\alpha}}{dt} = \sum\limits_\beta I_\beta P_{\beta\alpha}.
\end{equation}
Summation is carried out over three active flavours and expressions for
$P_{\beta\alpha}$ are listed in Appendix~\ref{app:neutrino-oscillations}
(Eqs.~\ref{eq:oscillation-probabilities}).

\subsection{The impact of sterile neutrinos}
\label{sec:sterile-dynamics}

As already mentioned, sterile neutrinos interact much more feebly than active
neutrinos do. Nevertheless, at some high temperature sterile neutrinos may
enter thermal equilibrium. Whether this happens or not depends on the thermal
history of the Universe before the onset of the synthesis.\footnote{For
  example in the \numsm model at early times ($T\gg 100\gev$) initial
  densities of sterile neutrinos are negligible~\cite{Bezrukov:08a}. Then the
  neutrinos come into equilibrium at temperature $T_+$ (typically $T_+ = 10
  \div 100$~GeV) and freeze-out at temperatures $T_-\sim
  0.5-5\gev$~\cite{Shaposhnikov:08a}.} Even if they were in thermal
equilibrium at early times, sterile neutrinos then necessarily decouple at
temperatures \emph{higher} than those of active neutrino decoupling.  If
sterile neutrinos were light and stable (or very long-lived), they would be
relativistic and propagate freely in the medium, yielding $N_{\rm eff}\approx
3+g_S $ together with active neutrinos ($g_S$ is the number of sterile neutrinos).  However, sterile neutrinos decay into
active neutrinos and other particles. The energies of the decay products may
be very different from the typical energies of plasma particles. For particles
that equilibrate quickly (such as electrons or photons), this ``injection''
results in the fast redistribution of the energy between all particles in
equilibrium and effectively the process looks like a temperature increase
(more precisely, it just slows down the cooling of the Universe).  But for
particles that either are not in equilibrium or are about to fall out of it,
such as active neutrinos at few MeV, the ``injection'' \textit{modifies} the
form of their spectra. The other mass-induced effect is that sterile neutrinos
may switch from the relativistic regime (when their average momentum is larger
than mass), that is established at large temperatures, to the non-relativistic
one, due to the gravitational redshift.

For the quantitative description of sterile neutrino dynamics we utilize the
Boltzmann equation similar to (\ref{eq:Boltzmann-equation}), replacing active
neutrino everywhere therein by sterile neutrino $\nu_S$
\begin{equation}
\label{eq:Boltzmann-equation-sterile}
\frac{df_S}{dt} = I_S
\end{equation}
Reactions contributing to the right-hand side together with their
probabilities are listed in Tables
\ref{tab:sterile-scatterings}--\vref{tab:sterile-decays} of
Appendix~\ref{app:matrix-elements}. Note that we neglect the processes
involving baryonic particles. However, they become important for temperatures
near the QCD crossover temperature $T_{QCD}\simeq 200\MeV$, when their density
is not negligible anymore. More scattering channels of sterile neutrino would
appear and their proper account is involved. However it seems to be reasonable
to assert that the only modification the account will bring is to \emph{lower} the decoupling temperature of sterile neutrinos.

Oscillation phenomenon does not affect significantly  sterile neutrinos and therefore Boltzmann equation in its original form (\ref{eq:Boltzmann-equation-sterile}) is still valid, contrary to what we have found out for active neutrinos. An argument in favor of this statement is explained in Appendix \ref{app:neutrino-oscillations}.  

When sterile neutrino is heavier than muon, the former particle can appear in
the decay $\nu_S \to \mu^- + e^+ + \bar\nu_e$. However, the branching fraction of
this decay mode does not even reach a percent for masses of sterile neutrino
we consider (see e.g.~\cite{Gorbunov:07a}). Therefore we can neglect influence
of both muons and other particles, appearing in the decay.

As a result we have six equations
(\ref{eq:Friedmann-first}),~(\ref{eq:Friedmann-second}),~(\ref{eq:Boltzmann-equation-oscillations}),
and~(\ref{eq:Boltzmann-equation-sterile}) describing primordial plasma at
temperatures of interest. These equations contain six unknowns -- scale factor
$a(t)$, temperature $T(t)$ and four neutrino distribution
functions, $f_{\nu_\alpha}$ and $f_S$. The system of equations is therefore
closed and we have solved it numerically at the \textsc{step i}.

\subsection{Course of nuclear reactions}
\label{sec:nuclear-dynamics}

Outcome of the nuclear reaction chains is found numerically. For the Standard
BBN model one of the earlier attempts was made with the code written by
L. Kawano~\cite{Kawano:1988vh,Kawano:1992ua}. However, the program in its
original form is inappropriate for the account of the BSM physics, and we
modified it for this work. Two technical remarks are in order here. First, we
used the 1992 version of the program~\cite{Kawano:1992ua} as a starting point,
and not the 1988 one,~\cite{Kawano:1988vh}. Therefore, the integration time
steps were taken small enough, so that the integration procedure did not
introduce an error, that was compensated as a shift in the resulting value of
the $Y_p$,\footnote{We denote by $Y_p$ the \emph{mass fraction} of the $^4$He,
  that is a fraction of the total baryon mass stored in the form of Helium-4}
the so-called ``Kernan correction''~\cite{Kernan:1994je}. Second, the code did
not take into account the Coulomb and the nucleon finite-mass corrections to
weak interaction rates, as well as radiative and finite-temperature
effects.\footnote{For the accurate account of these corrections, see e.g.\
  \cite{Iocco:08,Boyd:2010kj,Fuller:2010un,Coc:2011az}.} 
We do not calculate directly these effects, 
  but assume their net result to be in the form of the additive correction,
  which we took to be $\Delta Y_p = -0.0003$~\cite{Sarkar:1995dd}.
The tests described in
  Appendix~\ref{sec:stand-nucl-model} demonstrate an agreement of thus
  modified ``Kawano code'' with the results of the other code,
PArthENoPE~\cite{Pisanti:2007hk}, that takes a proper account of these
effects.

Presence of sterile neutrinos alters the standard dynamics of the temperature
and the expansion rate as well as the rates of weak interactions involving
neutrons and protons. These quantities are known from the \textsc{step i}, so
we have implemented the import of these data. Together with the change of
$\Delta Y_p$ indicated above, it has lead to the code, that became an
essential tool of \textsc{step ii} in our approach. The computations of
nuclide evolution started from temperatures of several MeV, when the chemical
equilibrium ceases to hold, up to temperatures $T_{\rm Fin}$.

\subsection{Adopted values of abundances of the light nuclei} 
\label{sec:abundances}

The observables of the BBN are concentrations, or abundances, of light
nuclides dispersed in the cosmos. The most relevant abundance in our problem
is that of $^4$He, as it is sensitive to the expansion rate of the Universe at
MeV temperatures and neutrino distribution functions.  The presence of sterile
neutrinos in plasma typically increases the concentration of $^4$He, described
by $Y_p$. 
 Accurate calculations carried out
  in the Standard Model \cite{Pisanti:2007hk} predict the values
\begin{align}
  \label{eq:3}
 Y_p^{\textsc{sbbn}} & = 0.2480  &(\tau_n = 885.7\text{ sec})\\
 Y_p^{\textsc{sbbn}} & = 0.2465 &(\tau_n = 878.5 \text{ sec})%
\end{align}
depending on the lifetime of neutron, $\tau_n$, see below.

There are two main methods of experimental determination of primordial Helium
abundance. The first one is related to the studies of low-metallicity
astrophysical environments and extrapolating them to zero metallicity case.
The $Y_p$ measurements are known to be dominated by systematic
uncertainties. Therefore we adopt the $Y_p$ values from the two most recent
studies, Refs.~\cite{Izotov:2010ca,Aver:2011bw} that have slightly different
implications. For recent discussion of various systematic uncertainties in
$^4$He determination, see~\cite{Mangano:11}.

In Ref. \cite{Izotov:2010ca} the value $Y_p = 0.2565 \pm 0.0010(\text{stat.})
\pm 0.0050(\text{syst.})$ was obtained. Therefore, the $2\sigma$ intervals
that we adopt in our studies are\footnote{We add the systematic errors
  linearly}
\begin{equation}
  \label{eq:6}
  Y_p = 0.2495-0.2635  \quad (\mbox{Ref.~\cite{Izotov:2010ca}}, 2\sigma~\text{interval})
\end{equation}
One notices that this result is more than $2\sigma$ away from the Standard
Model BBN predicted value of $Y_p$, Eq.~(\ref{eq:3}).

Using a subsample of the same data of~\cite{Izotov:2010ca}, a different group
had independently determined $Y_p$~\cite{Aver:2011bw}. From their studies we
adopt\footnote{We use the average value over metallicities, $\langle
  Y_p\rangle $ (Eq.~(8.2) of~\cite{Aver:2011bw}) and leave the systematic
  error from~\cite{Izotov:2010ca}.}  $Y_p = 0.2574 \pm 0.0036 (\text{stat.})
\pm 0.0050 (\text{syst.})$. As a result,
\begin{equation}
  \label{eq:7}
  Y_p =   0.2452-0.2696 \quad (\mbox{Ref.~\cite{Aver:2011bw}},
  2\sigma~\text{interval}) 
\end{equation}
(this values of $Y_p$ coincide with the Standard BBN one,~(\ref{eq:3}), at
about $1\sigma$ level).\footnote{ A study of~\cite{Peimbert:2007vm}, based
  on the independent dataset, provides the value $Y_p=0.2477\pm 0.0029$.
  Its upper bound becomes very close to that of~(\ref{eq:7}) if one employs an
  additional systematic uncertainty at the  level $\Delta Y_\text{syst} =
  0.010$ (twice the value of systematic  uncertainty
  of~\cite{Izotov:2010ca}).}

Second method of determination of Helium abundance is based on the CMB
measurements. This method is believed to determine truly pristine value of $Y_p$, not prone
to the systematics of astrophysical methods. However currently its
uncertainties are still much larger than of the first method. The present
measurements put it at
\begin{equation}
  Y_p = 0.22 - 0.40,\quad \Neff=3 \quad(\mbox{Refs.~\cite{Dunkley:2010ge,WMAP7}},
  2\sigma~\text{interval})
\label{eq:Yp-CMB-Neff3} 
\end{equation}
again consistent with the Standard Model BBN at $1.5\sigma$.
Here $\Neff$ is the so-called effective number of neutrino species
\begin{equation}
\label{eq:Neff-definition}
N_{\rm eff} = \frac{120}{7\pi^2} \frac{\rho_{\nu_e}+\rho_{\nu_\mu}+\rho_{\nu_\tau}}{T^4},
\end{equation} 
proportional to the ratio of the total energy, deposited
into the active neutrino species to that of photons. Notice, that the bound~(\ref{eq:Yp-CMB-Neff3}) is based on assumption that before the onset of the
  recombination epoch the effective number of neutrino species is close to its
  SM value $\Neff\approx 3$. As we will
  see later, sterile neutrinos can
  significantly distort $\Neff$. For the values of $\Neff$ strongly deviating
  from $3$ the CMB bounds on $Y_p$ gets modified. For example, the analysis carried out in~\cite{Dunkley:2010ge} reveals that
\begin{equation}
Y_p = 0.10-0.33,\quad \Neff=6 \quad (\mbox{Ref.~\cite{Dunkley:2010ge}},
  2\sigma~\text{interval}).
\label{eq:Yp-CMB-Neff6}  
\end{equation}
 The similar conclusion is reached if one employs the data of~\cite{Keisler:2011aw}.

The other element produced during the BBN is the Deuterium, and recent
  observations determine its abundance to be
\begin{equation}
  D/H = (2.2-3.5) \times 10^{-5} \quad (\mbox{Ref.~\cite{Iocco:08}},3\sigma~\text{interval}).
\label{eq:Deuterium-observations}
\end{equation}
This value is sensitive both to the   baryon-to-photon ratio and to \Neff.
In this work we adjust the value of baryon-to-photon
ratio $\eta$ at the beginning of the computation so that by $T_{\rm Fin} \sim
10\kev$ it is equal to the value given by cosmic microwave background
measurements \cite{WMAP7}.

Finally, we mention another important uncertainty originating from the
particle physics. There are two different measurements of neutron lifetime
$\tau_n$ that are at tension with each other. Particle Data Group
\cite{Nakamura:2010zzi} provides $\tau_n = 885.7 \pm 0.8\sec$, while
measurements performed by Serebrov et al. \cite{Serebrov:2010sg} result in
$\tau_n =878.5 \pm 0.8\sec $. We employ both results and explore the
differences they lead to in what follows.

\section{Results}
\label{sec:results}

\begin{figure*}[!t]
  \subfloat[$2\sigma$ upper bounds on sterile neutrino lifetime, based on
  different measurements of $Y_p$: Ref.~\protect\cite{Izotov:2010ca} (``Izotov \&
  Thuan''); Ref.~\protect\cite{Aver:2011bw} (``Aver et al.'');
  Refs.~\protect\cite{Dunkley:2010ge,WMAP7} (``CMB
  bound'')]{\label{fig:astro-bounds}\includegraphics[width=0.45\textwidth]{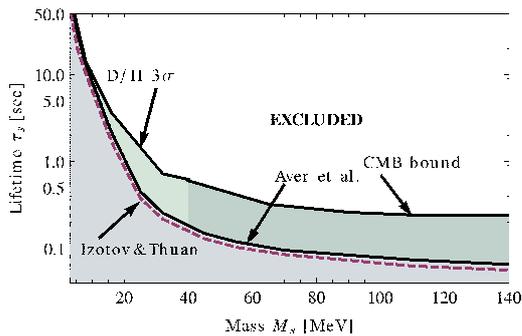}}
  \hspace{3em} \subfloat[Upper \emph{and lower} bounds on sterile neutrino
  lifetime, based on the measurements of~\protect\cite{Izotov:2010ca}. The upper curve
  is the same as the dashed curve in the left panel. ]{
    \includegraphics[width=0.45\textwidth]{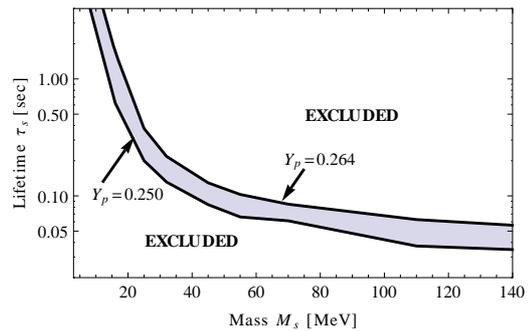}}
  \caption{Bounds (at $2\sigma$ level) on sterile neutrino lifetime as a
    function of their mass for various measurements of $Y_p$ (summarized in
    Section~\ref{sec:abundances}). All results are for mixing of sterile
    neutrino with electron flavour only (the dependence on the particular
    mixing pattern is very weak, see below). For the CMB bound, we present only
    the result for masses $M_s > \MassThresh$ where $\Neff \approx 3$.  For
    smaller masses we plot instead bounds based on $3\sigma$ Deuterium upper
    bound~\protect(\ref{eq:Deuterium-observations}).  For details,
    see Sec.~\protect\ref{sec:results} and Fig.~\protect\ref{fig:neff}.}
  \label{fig:izotov}
\end{figure*}

In this Section we present our main results: the bounds on sterile neutrino
lifetime as a function of their masses and mixing patterns, as well as the
bounds on the mixing angles. As discussed in the previous Section, there are
several systematic uncertainties in the determination of the $^4$He abundance
and therefore the results will depend on the adopted values of $Y_p$ (together
with the neutron lifetime, $\tau_n$). We summarize these systematic effects
below.

We start with comparing the upper bounds on sterile neutrino lifetime for
different values of $Y_p$ (see Section~\ref{sec:abundances}). The
Fig.~\ref{fig:astro-bounds} shows that the bounds from the two recent
works~\cite{Izotov:2010ca,Aver:2011bw} are quite similar (the difference is of
the order of 30\%). The bound, based on~\cite{Peimbert:2007vm} would
  give a result, similar to~\cite{Aver:2011bw} as discussed above.

\begin{figure}[!tp]
  \centering
\includegraphics[width=.45\textwidth]{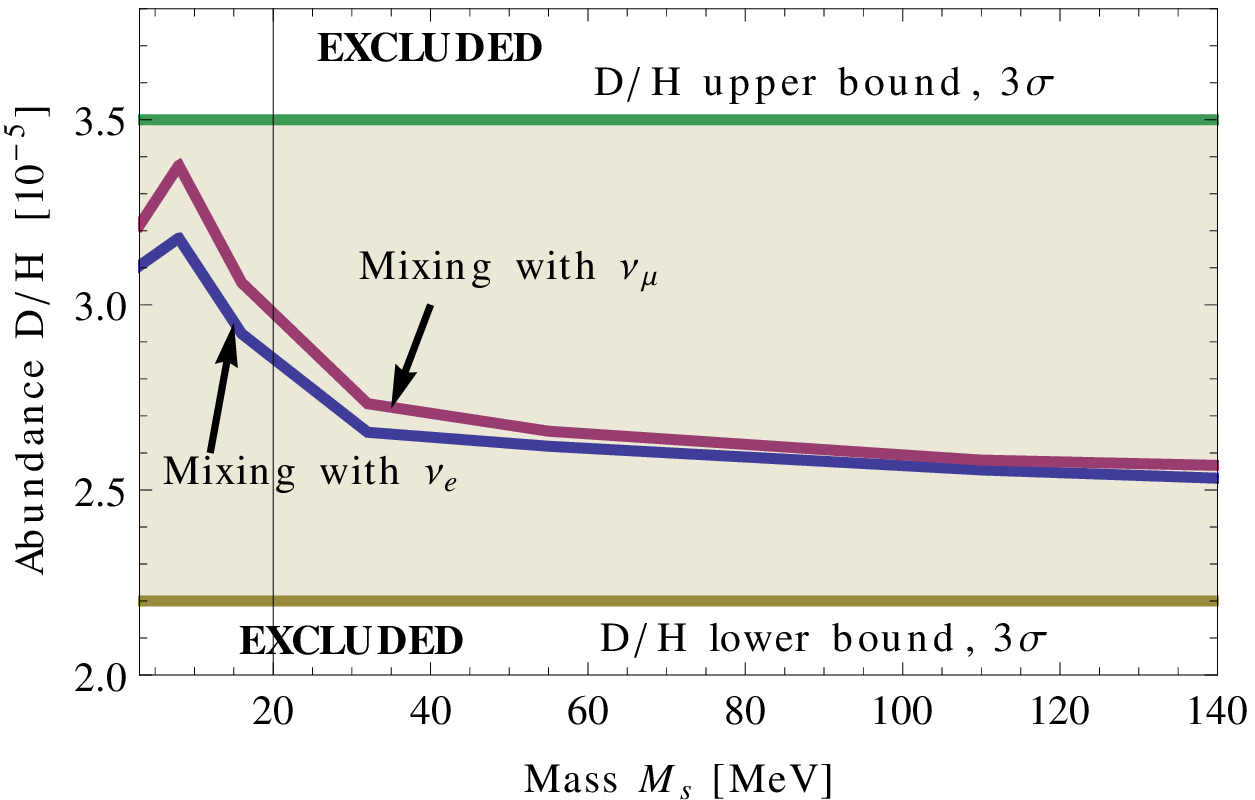}~
  \includegraphics[width=.45\textwidth]{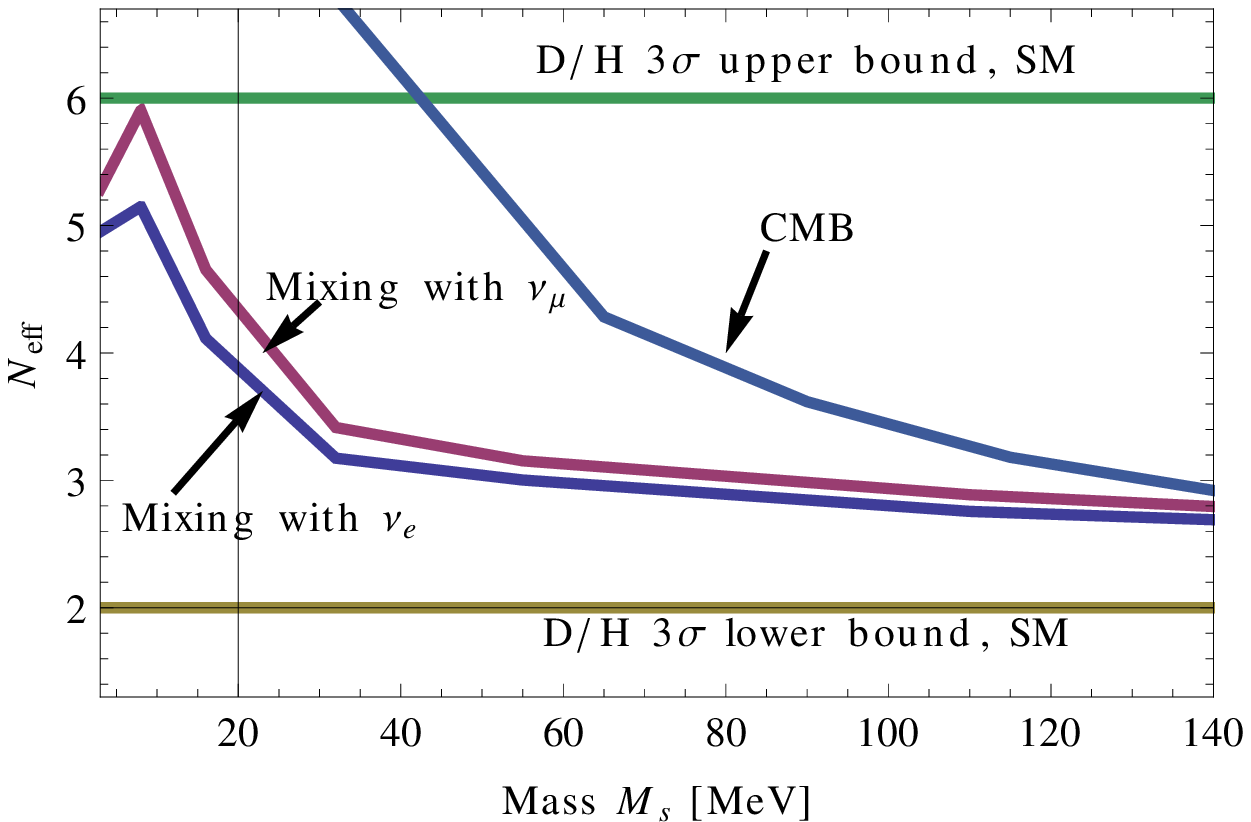}
  \caption{\textbf{Left:} Deuterium abundance, with the shaded region
    corresponding to the \emph{allowed} $3\sigma$ range, based
    on~\protect\cite{Iocco:08}. \textbf{Right:} Effective number of
    neutrino species
    (the ratio of the effective neutrino temperature to the photon temperature
    at $T\sim \text{few}\kev$) as a result of decay of sterile neutrino. The
    horizontal ``SM'' lines indicate $\Neff$ that corresponds to the boundary
    of the $3\sigma$ range~\protect\cite{Iocco:08}, in the SM with the number of
    relativistic species deviating from $\Neff\approx 3$.  In both panels,
    parameters of sterile neutrinos correspond to the upper bound on $Y_p$
    from \protect\cite{Izotov:2010ca} (see Eq.~(\protect\ref{eq:6})), except of the ``CMB'' line that corresponds to the upper bound from~\protect\cite{Dunkley:2010ge,WMAP7} (see Eq.~(\protect\ref{eq:Yp-CMB-Neff3})).
}
  \label{fig:neff}
\end{figure}

  For the CMB bound in Fig.~\ref{fig:astro-bounds}, we present only the
  results for masses $M_s > \MassThresh$ where $\Neff$ does not deviate
  significantly from $3$. Fig.~\ref{fig:neff} indicates that for smaller masses the
  number of effective neutrino species increases significantly.  It in turn
  affects the CMB helium bounds (c.f.\ Eqs.~(\ref{eq:Yp-CMB-Neff3}) and
  (\ref{eq:Neff-definition})). The accurate
    account of this effect goes beyond the scope of this work and we choose
  instead to plot stronger deuterium-based bounds (those of
  Fig.~\ref{fig:neff}) in Fig.~\ref{fig:astro-bounds} for $M_s \lesssim
  \MassThresh$.

The lower bound on $Y_p$ from the recent work of~\cite{Izotov:2010ca} is above
the Standard BBN value~(\ref{eq:3}) at $\sim 2\sigma$ level (see
however~\cite{Mangano:11}). The presence of sterile neutrinos in plasma of
course relaxes this tension and therefore at $2\sigma$ the adopted values of
$Y_p$ (Eq.~\ref{eq:6}) provide both upper and \emph{lower} bounds on sterile
neutrino lifetime. This is shown in Fig.~\ref{fig:izotov}, right panel. At
$3\sigma$ level the measurements of~\cite{Izotov:2010ca} are consistent with
Standard BBN and the lower bound disappears.

Fig.~\ref{fig:neff} shows the changes in Deuterium abundance and in the
effective number of neutrino species, caused by sterile neutrinos (with
parameters corresponding to the upper bound based
on~\cite{Izotov:2010ca}). For these values of parameters the abundance lies
within the $3\sigma$ boundaries~(\ref{eq:Deuterium-observations}). And for the
highest effective number of neutrinos reached, $\Neff=6$, $D/H$ is close to
the $3\sigma$ upper bound. Notice that the same relation between \Neff and
$D/H$ is observed in the model without new particles but with the effective
number of neutrinos different from $3$. The effective number of neutrino
  species does not define the Helium abundance though. Otherwise the same
  $Y_p$ bound~\cite{Izotov:2010ca} would predict \emph{only one particular}
  value of \Neff, which is not case, as the inspection of Fig.~\ref{fig:neff}
  shows.

The influence of another systematic uncertainty (the lifetime of neutron,
$\tau_n$) is negligible. Indeed, the relative difference between sterile
neutrino lifetimes were found to be of the order of $5\%$ for two choices of
$\tau_n$ -- from \cite{Serebrov:2010sg} and from \cite{Nakamura:2010zzi}
(taking the same $Y_p$ bound from~\cite{Aver:2011bw}).

\begin{figure}[!t]
  \centering%
  \subfloat[Bounds based on astrophysical measurements of $Y_p$
  of~\cite{Aver:2011bw}]%
  {\includegraphics[width=.42\textwidth]{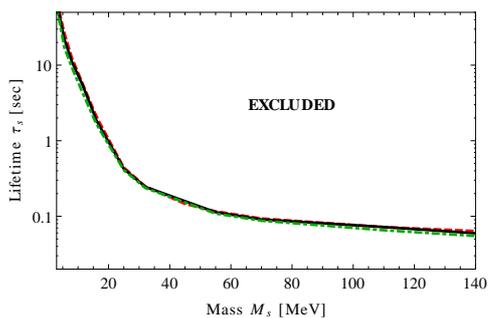}}
  \hspace{3em} \subfloat[Bounds from CMB (notice different $y$-axis range!).]
  {\label{fig:cmb-upper}\includegraphics[width=.42\textwidth]{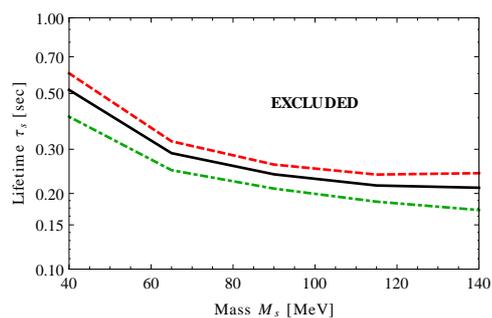}}
  \caption{Upper bound for sterile neutrino lifetime for different mixing
    patterns: mixing with $\nu_e$-only (red dashed line), $\nu_\mu$-only
    (green dashed-dotted line) and equal mixing with $\nu_e$ and $\nu_\mu$
    flavours (black solid line). All bounds are derived for the lifetime of
    neutron $\tau_n$ adopted from~\protect\cite{Nakamura:2010zzi}.  The effect
      of different mixing patterns is at the level $\sim 10-50\%$
 and can only
      be seen in the right panel because of the different $y$ axis.  In the
    right panel, only the masses $M_s > \MassThresh$ are presented. For
    details, see Sec.~\protect\ref{sec:results} and Fig.~\protect\ref{fig:neff}. }
  \label{fig:mixing-patterns}
\end{figure}

\begin{figure}[!t]
  \centering%
  \subfloat[Bounds based on astrophysical measurements of $Y_p$ of~\protect\cite{Aver:2011bw}]%
  {\includegraphics[width=.42\textwidth]{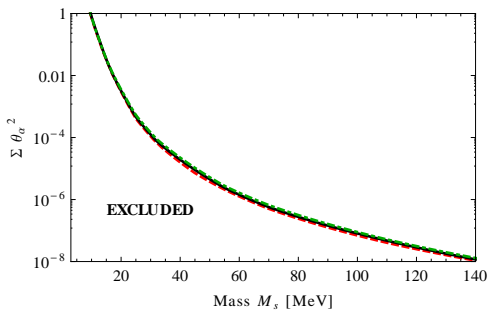}}
  \hspace{3em} \subfloat[Bounds from CMB]
  {\label{fig:cmb-lower}\includegraphics[width=.42\textwidth]{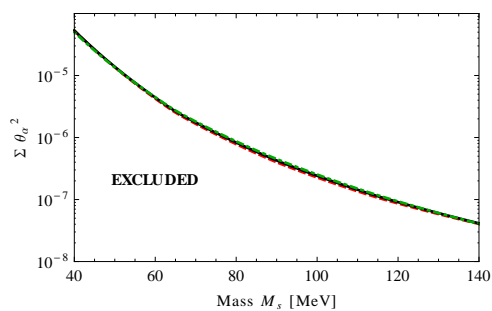}}
  \caption{\emph{Lower} bound on mixing angles of sterile neutrinos for different mixing
    patterns: mixing with $\nu_e$-only (red dashed line), $\nu_\mu$-only
    (green dashed-dotted line) and equal mixing with $\nu_e$ and $\nu_\mu$
    flavours (black solid line). Both types of bounds are derived by assuming lifetime
    of the neutron $\tau_n$ from~\protect\cite{Nakamura:2010zzi}.  In the right panel, only the masses $M_s > \MassThresh$ are presented. For details, see Sec.~\protect\ref{sec:results} and Fig.~\protect\ref{fig:neff}.}
  \label{fig:mixing-bounds}
\end{figure}

Next we investigate the dependence of the resulting bounds on the mixing
patterns of sterile neutrinos. Naively, one would expect that sterile
neutrinos mixing ``only with $\nu_e$'' and ``only with $\nu_\mu$'' should have
different effect of $Y_p$.  However, it is the energy ``injection'' rate
(i.e.\ the overall decay rate of sterile neutrinos) that is \emph{more
  important} for the dynamics of plasma before the onset of
nucleosynthesis. This quantity depends on the lifetime $\tau_s$ and the mass
$M_s$ of the neutrino. Mixing patterns affect mostly the concentration of
particular decay products, but not the injection rate. In addition, the
neutrino oscillations (fast at the BBN epoch) make the difference between
flavours less pronounced (see Appendix~\ref{app:neutrino-oscillations}). As a
result, mixing patterns give essentially the same results with the difference
at the level of tens of per cent (see
Figs.~\ref{fig:mixing-patterns},~\ref{fig:mixing-bounds}).

\section{Discussion}
\label{sec:discussion}
In this work we considered the influence of decaying particles with the masses
few MeV --140 MeV on the primordial abundance of light elements ($D$ and
$^4$He). Such particles appear in many cosmological scenarios
\cite{Shaposhnikov:08a,Canetti:10a,Fuller:2011qy,Dolgov:1997it,Dolgov:88,Kawasaki:1993gz,Kawasaki:2000en,Hannestad:04,Gelmini:04,Gelmini:2009xd,Fuller:2009zz}. Particularly,
we concentrated on the properties of sterile neutrinos and derived constraints
on their lifetime imposed by the present measurements of primordial Helium
abundance $Y_p$.  Sterile neutrinos are super-weakly interacting particles,
quadratically mixed with the active flavours.

We analyzed the case of one Majorana sterile neutrino with 4 degrees of
freedom (if sterile neutrinos were kept in thermal equilibrium it would be
equivalent to $g_s = 2$ species of active neutrinos). Since the plasma
evolution is mostly affected by the overall decay rate of sterile neutrinos,
the lifetime bounds that we obtained are essentially independent of the
particular mixing patterns, as
  Figs.~\ref{fig:mixing-patterns},\ref{fig:mixing-bounds} demonstrate.

\begin{figure}[!t]
  \centering
  \includegraphics[width=0.45\textwidth]{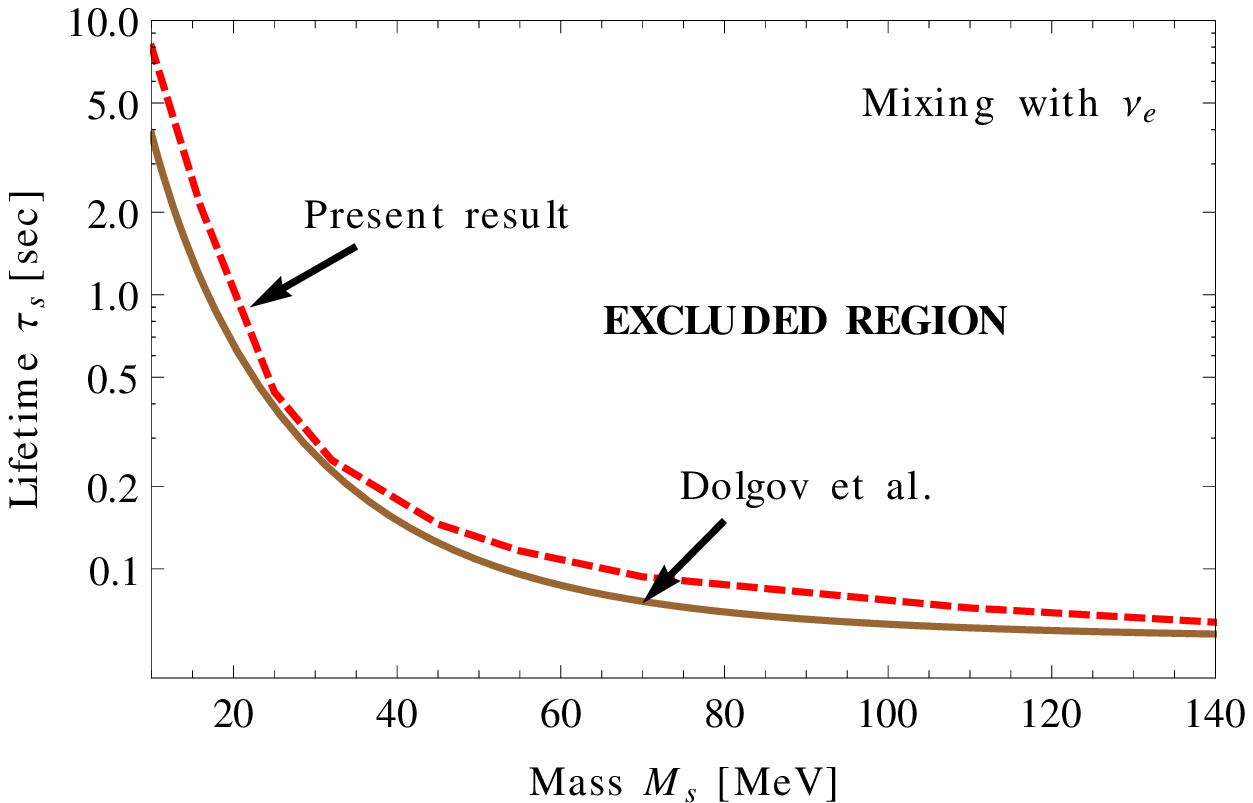}
  \includegraphics[width=0.45\textwidth]{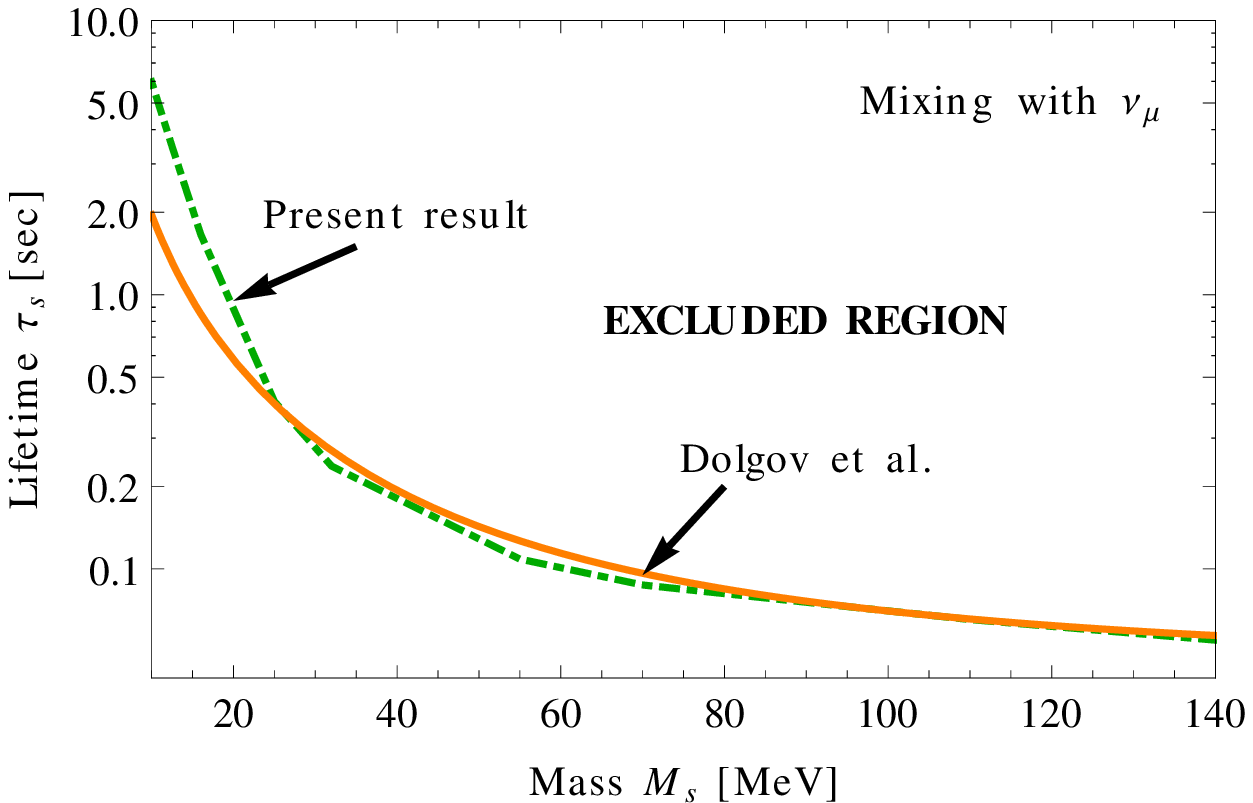}
  \caption{Comparison with the previous results of~\protect\cite{Dolgov:00a,Dolgov:00b}}
  \label{fig:dolgov}
\end{figure}

In the paper \cite{Dolgov:00b} a similar model was considered with one
\emph{Dirac} sterile neutrino. Dirac sterile neutrino has the same 4 degrees
of freedom and influences primordial plasma in the same way (if it has the
same spectrum, lifetime and mixing pattern). However, in \cite{Dolgov:00b}
effect of active-neutrino oscillations was not taken into account, and some
simplifying approximations like Boltzmann statistics were employed. To provide
corresponding analysis we wrote code that solves more accurate Boltzmann
equations describing kinetics of neutrino than what were used in
\cite{Dolgov:00b}.  We compare the results of this work with the previous
bounds~\cite{Dolgov:00a,Dolgov:00b} in Fig.~\ref{fig:dolgov}. We see
  that our results are broadly consistent with the previous works. The
  differences for a given mixing pattern of sterile neutrinos can be as large
  as a factor of 2.5 for some masses.

The presence of sterile neutrinos in the plasma affects the effective number
of neutrino degrees of freedom, \Neff. Fig.~\ref{fig:neff}, right panel shows
that \Neff between $2.7$ and $6$ are possible for different mixing
angles and masses, which could explain a larger than 3 values of $N_{\rm
  eff}$, reported recently in several CMB observations (see
e.g.~\cite{Dunkley:2010ge,Benson:2011ut,Keisler:2011aw}, but
also~\cite{Moresco:2012by}).

Decaying sterile neutrinos with the masses $100-500 \mev$ and lifetimes from
seconds to minutes and their influence on $N_{\rm eff}$ and entropy production
have been recently considered in~\cite{Fuller:2011qy} (see
also~\cite{Asaka:06}) where it was demonstrated that they can lead to $N_{\rm
  eff}\neq 3$ and can therefore be probed with the CMB measurements. The
results of the present work demonstrate that in the region $100-140\mev$ where
we overlap with the parameter space, studied in~\cite{Fuller:2011qy}, the
primordial nucleosynthesis restricts the lifetime of sterile neutrinos to be
well below 1~sec (see Fig.~\ref{fig:izotov}, left panel).

\begin{figure}[!t]
\centering
\includegraphics[width=.49\textwidth]{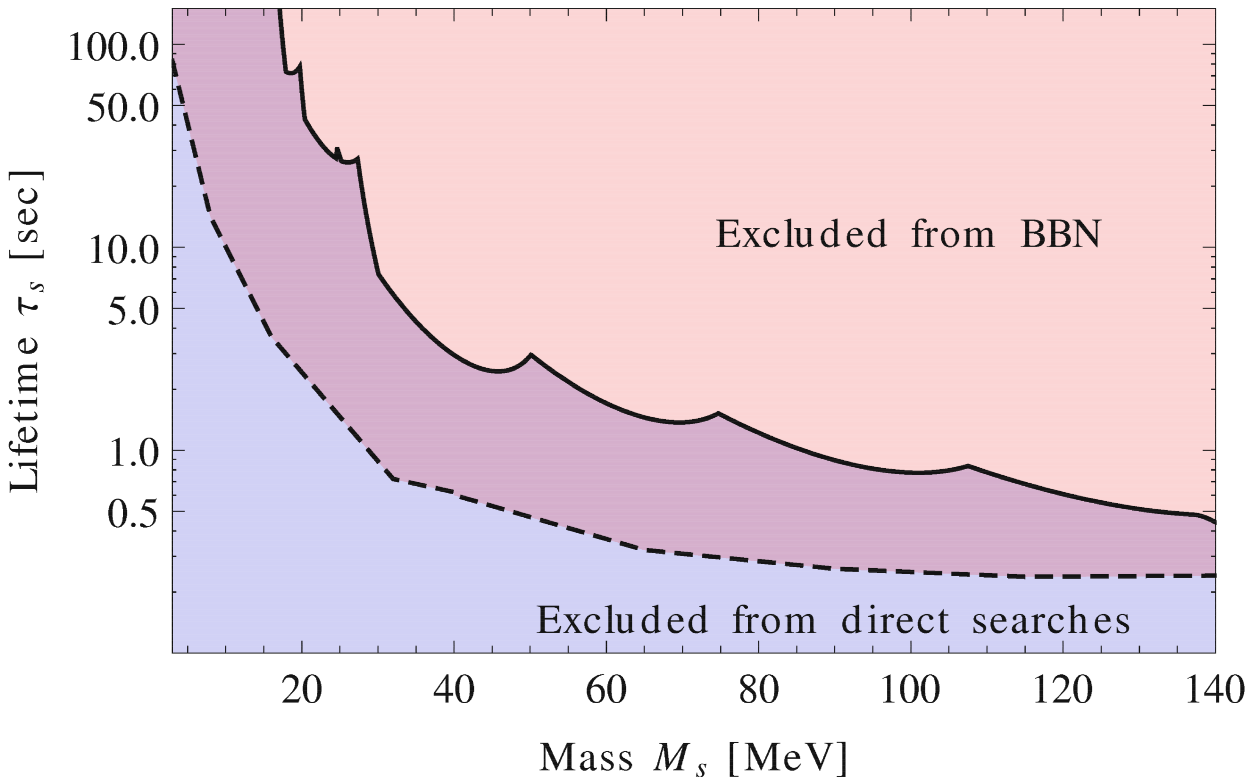}~
\includegraphics[width=.49\textwidth]{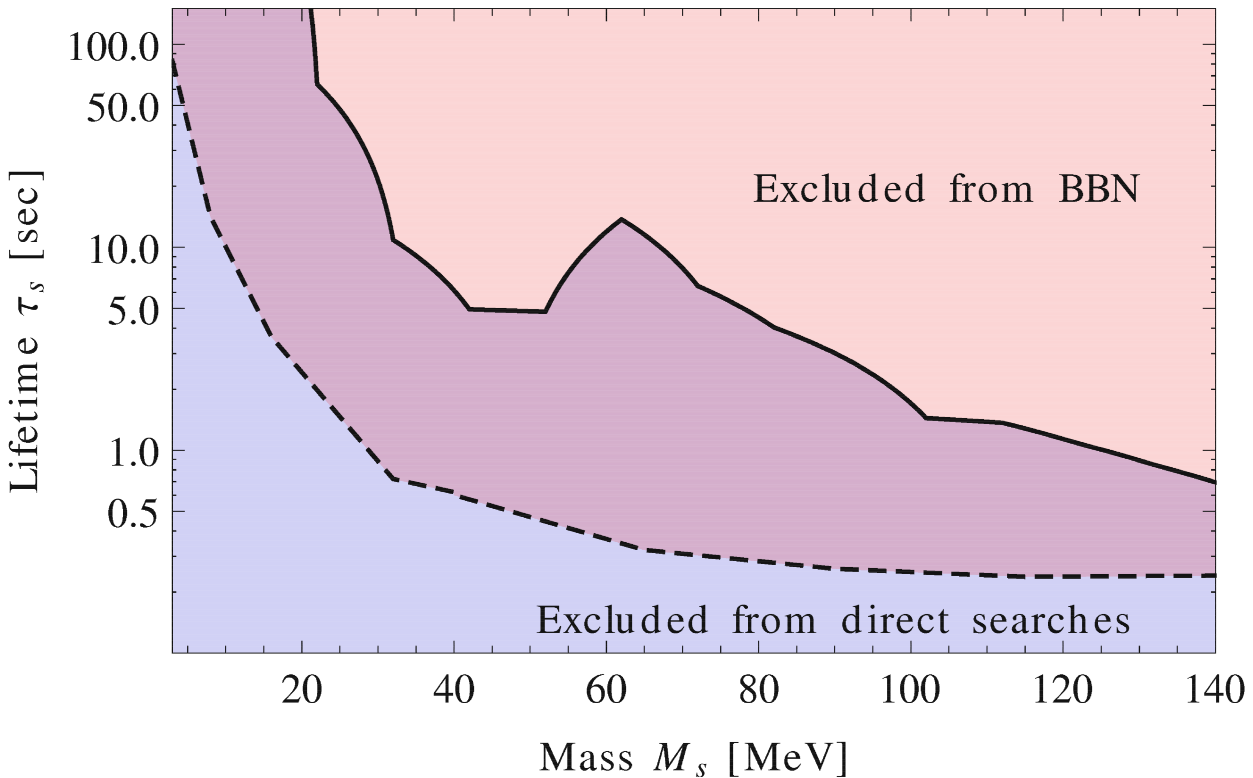}
\caption{Experimental $3\sigma$ \emph{lower} bounds on the lifetime of sterile
  neutrinos \protect\cite{Ruchayskiy:2011aa} (solid line), combined with the
  \emph{upper} bounds from this work (dashed line), corresponding to the
  \emph{weakest} bound in Fig.~\protect\ref{fig:astro-bounds}.  The
    accelerator bounds are for two Majorana sterile neutrinos solely
    responsible for neutrino oscillations.  \textbf{Left:} normal hierarchy,
  \textbf{right:} inverted hierarchy. Combination of BBN bounds with
    direct experimental searches demonstrates that sterile neutrinos with the
    masses in 1-140 MeV range, solely responsible for neutrino oscillations
    are ruled out.  See
  Secs.~\protect\ref{sec:results},\protect\ref{sec:discussion} for details.}
\label{fig:lifetime-BBN}
\end{figure}

\begin{figure}[!t]
  \centering
  \includegraphics[width=.6\textwidth]{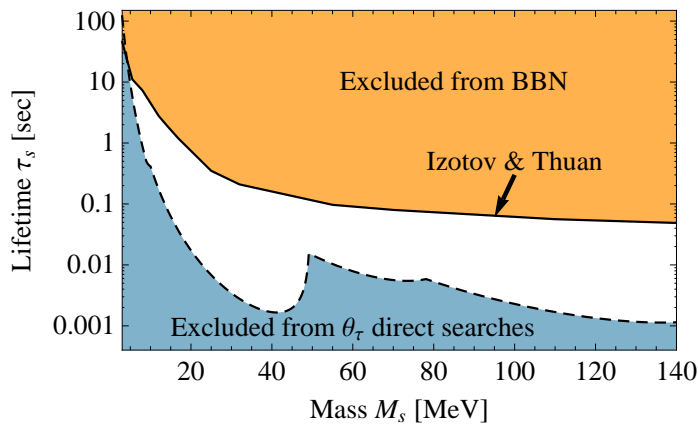} %
  \caption{Comparison of direct accelerator constraints and BBN bounds,
    based on the Helium-4 measurements of~\cite{Izotov:2010ca} in the model
    where sterile neutrinos mix with $\nu_\tau$ \emph{only}. Unlike the case,
    presented in Fig.~\protect\ref{fig:lifetime-BBN} there is an allowed
    region of parameter space for most of the masses below 140~MeV.}
  \label{fig:lifetime-tau}
\end{figure}

Finally, it is interesting to compare the upper bound on sterile neutrino
lifetime, derived in this paper with the \emph{lower bounds} that come from
direct experimental searches for sterile neutrinos
(see~\cite{Gorbunov:07a,Asaka:2011pb,Ruchayskiy:2011aa}). These latter bounds
are based on the assumption that sterile neutrinos with four degrees of
freedom are solely responsible for the observed pattern of neutrino
oscillations via the see-saw mechanism~\cite{Ruchayskiy:2011aa}. The
appropriate comparison, based on~\cite{Ruchayskiy:2011aa}, is presented in
Fig.~\ref{fig:lifetime-BBN}.  No allowed values of sterile neutrino lifetimes
for $1 \MeV \lesssim M_s < 140\MeV$ exist for either type of neutrino mass
hierarchy (i.e.\ the upper bound is \emph{smaller} than the lower bound, see
the purple double-shaded region in Fig.~\ref{fig:lifetime-BBN}). Notice, that
if the astrophysical bounds on Helium~\cite{Izotov:2010ca,Aver:2011bw} were
used for $M_s\gtrsim \MassThresh$ in Fig.~\ref{fig:lifetime-BBN}, instead of
the CMB bound, the resulting lifetime bounds would become stronger (by as much
as a factor of 4) in this mass range. \emph{We stress that for this
    conclusion it is essential} that MeV sterile neutrinos are responsible for
  neutrino oscillations.  For example, a model in which sterile neutrinos
  couple to $\nu_\tau$ \emph{only} (and therefore do not contribute to the
  mixing between active neutrino flavours), \emph{is allowed} even if one
  confronts the strongest BBN bounds (based on the astrophysical Helium
  measurements) with the direct accelerator bounds, see
  Fig.~\ref{fig:lifetime-tau} for details.

\subsubsection*{Acknowledgments}
\label{sec:acknowledgements}

We would like to thank A.~Boyarsky, D.~Gorbunov, S.~Hansen, D.~Semikoz,
M.~Shaposhnikov for valuable discussions and for help and encouragement during
various stages of this project. We specially thank D.~Semikoz for sharing with
us the original version of the code, used
in~\cite{Dolgov:00a,Dolgov:00b,Dolgov:98addendum,Dolgov:1997it} and J.~Racle
for writing an initial version of the BBN code as a part of his Master's
project \cite{Racle:08} at EPFL. A.I. is also grateful to S.~Vilchynskiy, Scientific and
Educational Centre of the Bogolyubov Institute for Theoretical Physics in
Kiev, Ukraine\footnote{\url{http://sec.bitp.kiev.ua}} and to Ukrainian Virtual
Roentgen and Gamma-Ray Observatory
VIRGO.UA.\footnote{\url{http://virgo.org.ua}}
The work of A.I. was supported in part from the Swiss-Eastern European cooperation project (SCOPES)
No.~IZ73Z0\_128040 of Swiss National Science Foundation. A.I. acknowledges support from the ERC Advanced Grant 2008109304.

 \appendix

\section{Tests of the numerical approach}
\label{sec:numerical-tests}

\begin{table}[!t]
  \centering
  \begin{tabular}[c]{l|c|c}
    Code & $Y_p$ for $\tau_n$ from PDG~\protect\cite{Nakamura:2010zzi} & 
    $Y_p$ for $\tau_n$ from \protect\cite{Serebrov:2010sg} \\
    \hline
    (Modified) Kawano code~\protect\cite{Kawano:1992ua} & 0.2472 & 0.2457 \\
    PArthENoPE code~\protect\cite{Pisanti:2007hk} & 0.2480 & 0.2465\\
    \hline
    Difference & -0.0008 & -0.0008
  \end{tabular}
  \caption{Values of Helium abundance $Y_p$ in the Standard Model BBN (SBBN)
    and their dependence on the neutron lifetime, $\tau_n$.  }
  \label{tab:yp}
\end{table}

The Section below summarizes the comparison of the present work with the
previous ones that analyzed the influence of the MeV particles on primordial
nucleosynthesis.  Throughout this Section, we normalize scale factor by
imposing condition $aT=1$ at the initial moment. Conformal momentum is $y=pa$
with the same normalization of the scale factor. In the figures that
  contain both the solid and the dashed curves, the former correspond to the
  results obtained with our code, and the latter -- to the original results of
  the other papers.

\subsection{Standard Model BBN}
\label{sec:stand-nucl-model}

\begin{figure}
\centering
  \includegraphics[width=.6\textwidth]{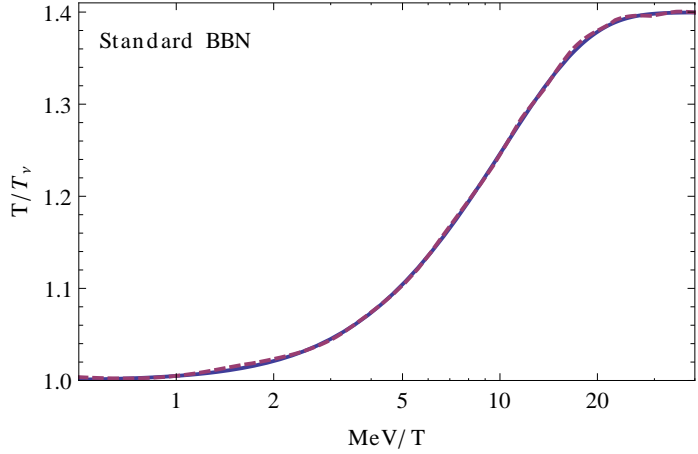}
  \caption{$T/T_\nu$ as a function of inverse temperature $T^{-1}$. The \textbf{solid} line is produced by the code of the present work, the \textbf{dashed} -- the result of~\protect\cite{Dolgov:97sm}.}
  \label{fig:SM-aT-evolution}
\end{figure}

\begin{figure}
\centering
 \includegraphics[width=.45\textwidth]{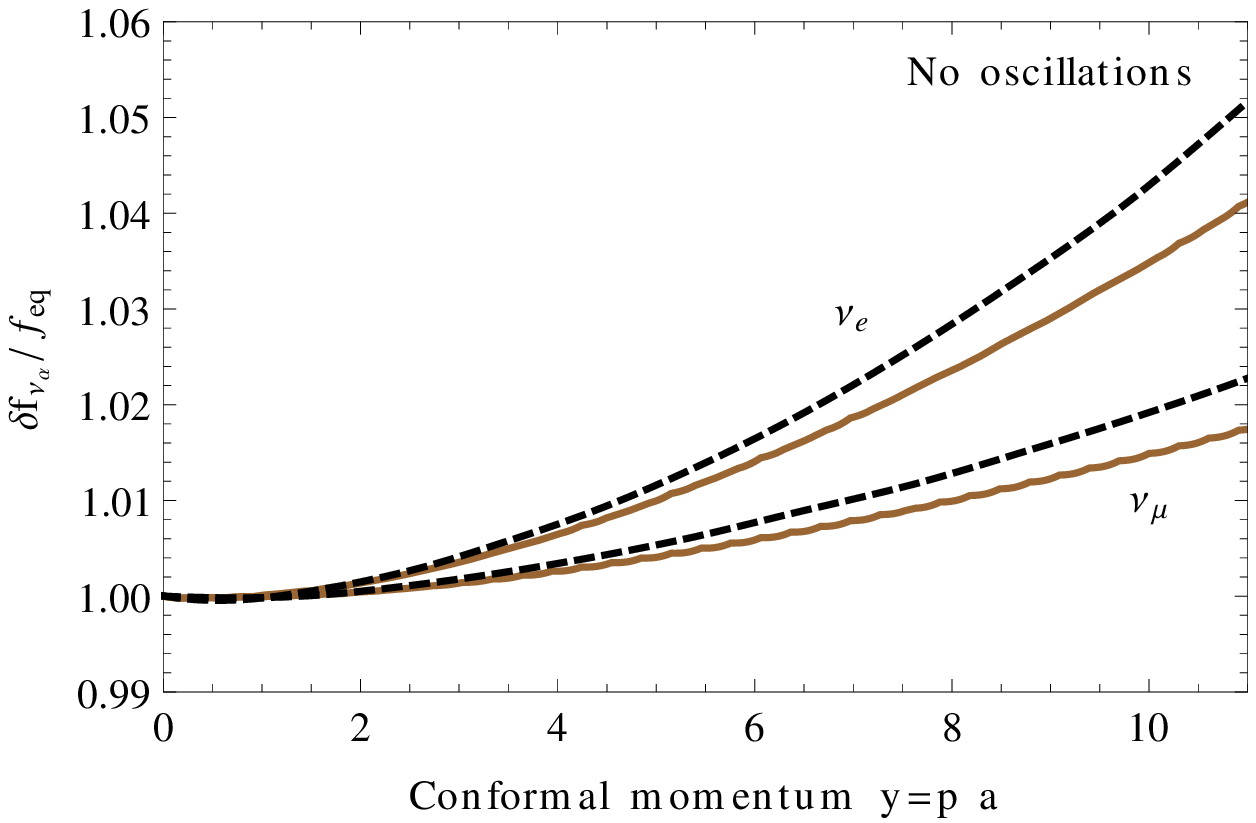}
\includegraphics[width=.45\textwidth]{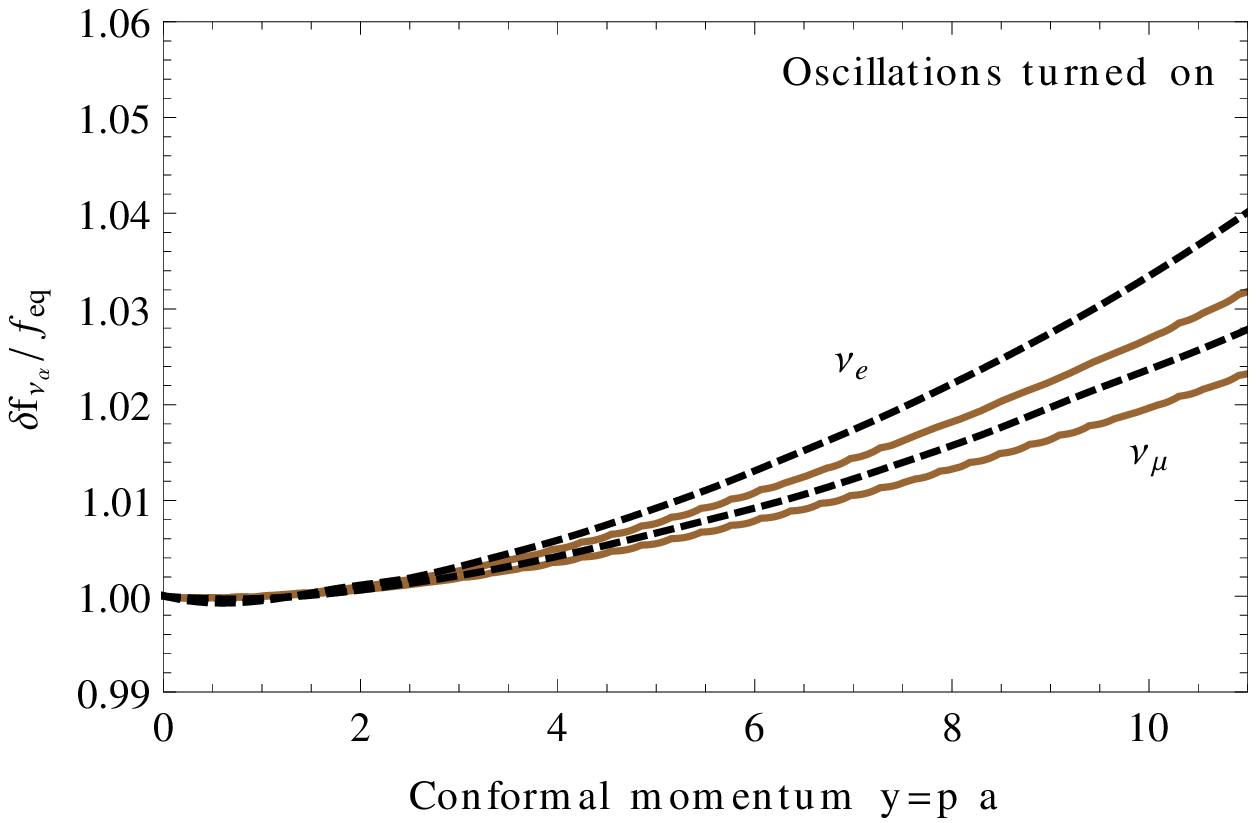}
\caption{ Relative distortions of neutrino spectra before the onset of
  BBN. \textbf{Left:} neutrino flavour oscillations are neglected,
    \textbf{right:} the oscillations are taken into account, with the
    parameter choice $\theta_{13}=0$, $\sin^2\theta_{23}=0.5$,
    $\sin^2\theta_{12}=0.3$ used in~\protect\cite{Mangano:2005cc}. In both panels,
  the pair of upper curves shows the distortion of the electron neutrino, the
  lower -- of $\nu_\mu$. In each pair, the solid curve is the result of this
  work, and the dashed is from Fig.~2 of~\protect\cite{Mangano:2005cc}.}
  \label{fig:final-spectra-SM}
\end{figure}

\begin{figure}
 \includegraphics[width=.5\textwidth]{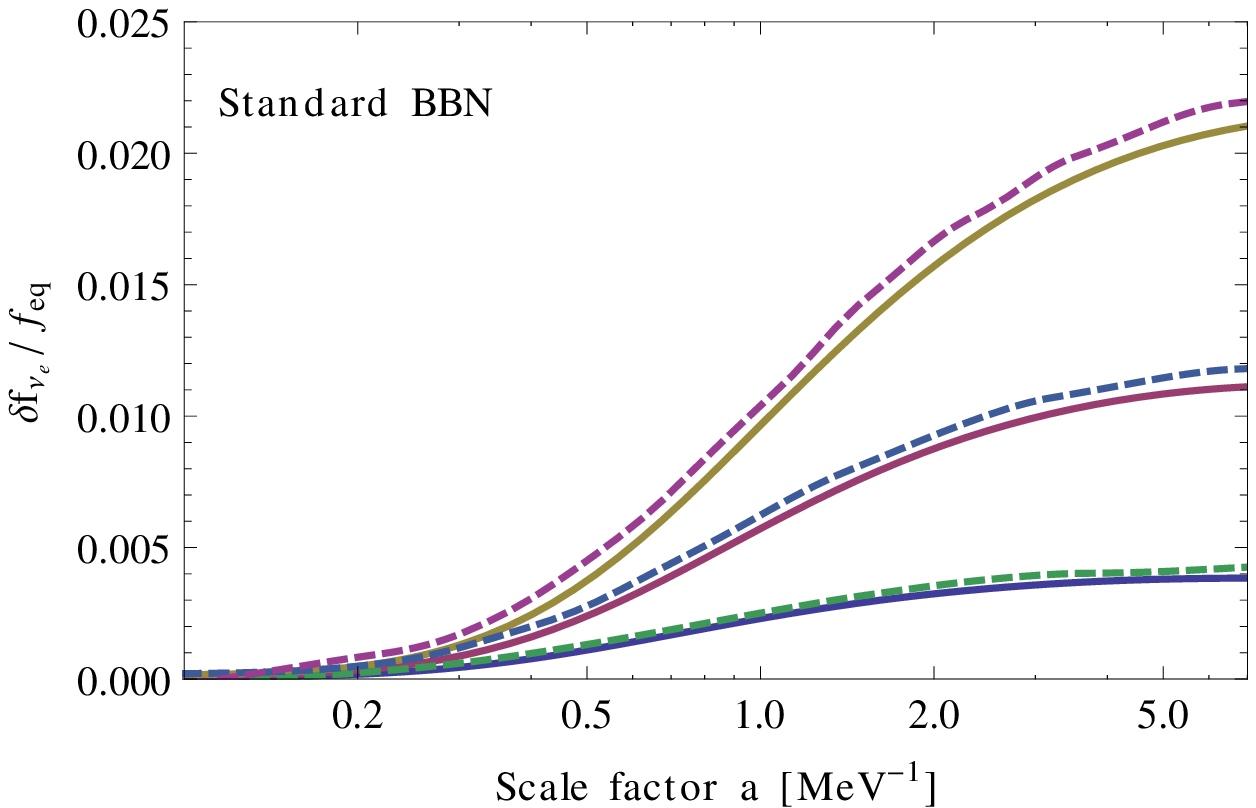}
 \includegraphics[width=.5\textwidth]{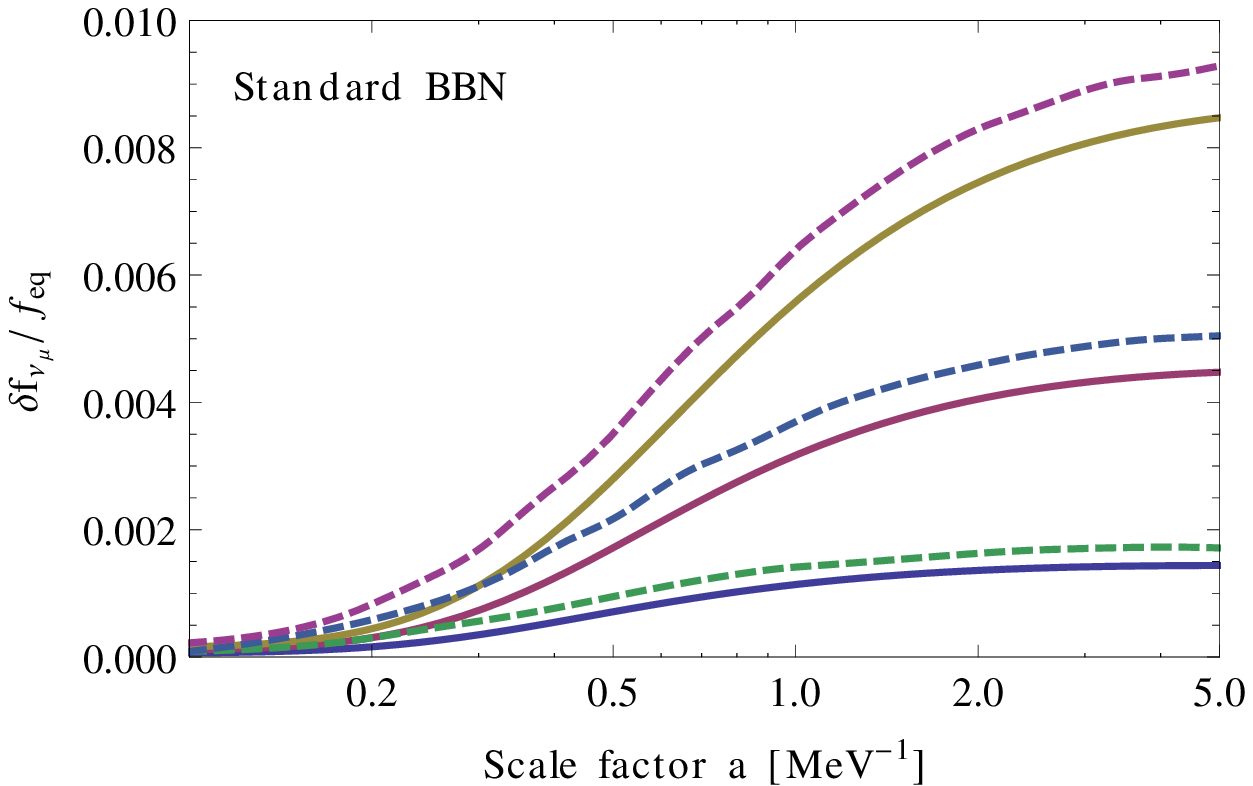}
  \caption{\textbf{Left:} Relative distortion of $\nu_e$ spectra $\delta f_{\nu_{e}} / f_{\rm eq}$ for conformal momenta $y=3,5,7$ (from bottom to up). \textbf{Right:} The same, but for muon neutrino. In each pair of curves the solid one corresponds to this work and the dashed one is from~\protect\cite{Dolgov:97sm}.}
  \label{fig:sm-deltaf}
\end{figure}

First we considered the nucleosynthesis in Universe filled with the Standard
Model particles only.  We compute the actual \emph{non-equilibrium} form of
the active neutrino spectra during their decoupling.  The results of the
present work are compared with those
of~\cite{Dolgov:97sm,Dolgov:98addendum,Mangano:2005cc}.
In~\cite{Dolgov:97sm,Dolgov:98addendum} neutrino oscillations were neglected,
while in~\cite{Mangano:2005cc} the effect was taken into
account. Fig.~\ref{fig:SM-aT-evolution} shows the evolution of the quantity
$aT$ as a function of temperature. It is identical to the Fig.~1 in
Ref.~\cite{Dolgov:97sm}.
Figures~\ref{fig:final-spectra-SM},\ref{fig:sm-deltaf} show how distorted
neutrino spectra $f_{\nu_\alpha}$ are, compared to the thermal distribution
$f^{eq}= (e^{y} + 1)^{-1}$. One can see good agreement between the results.
We believe that the difference, that is present nevertheless, arises solely
due to our one-step time integration method of the stiff kinetic equations,
that is not as accurate as the method employed in 
Refs.~\cite{Dolgov:98addendum,Mangano:2005cc}.

We turned off flavour oscillations and compared asymptotic values of ratio
$aT$ at low temperatures together with the effective number of neutrino
  species, $N_{eff}$. For the former quantity,
Refs.~\cite{Dolgov:98addendum,Mangano:2005cc} present values $1.3991$ 
  and $1.3990$, respectively. On the other hand, we derived $1.3996$.  
  For the number of neutrino species in absence of neutrino oscillations, the
  same Refs.~\cite{Dolgov:98addendum,Mangano:2005cc} provide numbers $3.034$
  and $3.035$, respectively, while we get $3.028$.\footnote{
  Ref.~\cite{Mangano:2005cc} the takes into account both the effects of
  neutrino oscillations and QED corrections, the latter changes the result
  significantly. As a result we could not compare the effect of \emph{neutrino
    oscillations only}.}

The resulting $Y_p$ is summarized in Table~\ref{tab:yp} for different values
of neutron lifetime $\tau_n$. We also provide a comparison of the modified
version of the Kawano code~\cite{Kawano:1992ua} that we adopted for computing
nuclear reactions with a newer code, PArthENoPE~\cite{Pisanti:2007hk}. By
comparing the results of PArthENoPE and the modified KAWANO code, we find the
former to be larger by 0.0008 than the latter.
  We use the shift $\Delta Y_p=-0.0008$ as a correction in our subsequent
  results.


\subsection{Test of energy conservation}
\label{sec:energy-conservation}
If all weak reactions involving electrons and positrons are turned off, neutrinos decouple from the rest of plasma. Then the energy conservation law (\ref{eq:Friedmann-second}) holds \textit{separately} for the neutrino component and for the remaining particles. In approximation of zero mass of electron we obtain
\begin{equation}
\label{eq:aT-conservation}
  \frac{d(a T)}{dt} = 0 
\end{equation}
similar to Eq. (\ref{eq:temperature-decay}). As a corollary, product $aT$ is
conserved. On the other hand, our code solves the equation
(\ref{eq:Friedmann-second}) involving \textit{all} medium components
simultaneously. And it turns out that the relation (\ref{eq:aT-conservation})
is not a trivial consequence of the numerical computation. Therefore the check
of the conservation serves as a test of the code. We considered separately
scattering and decay processes involving neutrinos and observed conservation
of $a T$ with precision of order $0.2\%$.

\subsection{Heavy sterile Dirac neutrino}

Next we have tested model with \textit{one} sterile \textit{Dirac} neutrino
$\nu_S$ with mass $M_s=33.9\MeV$, mixed with $\nu_\tau$~\cite{Dolgov:00a}. This
neutrino was assumed to be in thermal equilibrium with plasma at $T\gtrsim
50\MeV$. To simplify the problem, the authors of~\cite{Dolgov:00a} used the
\textit{Boltzmann equilibrium} statistics for active species in collision
integral for a sterile neutrino.

\begin{figure}[tp]
  \centering
  \includegraphics[width=0.6\textwidth]{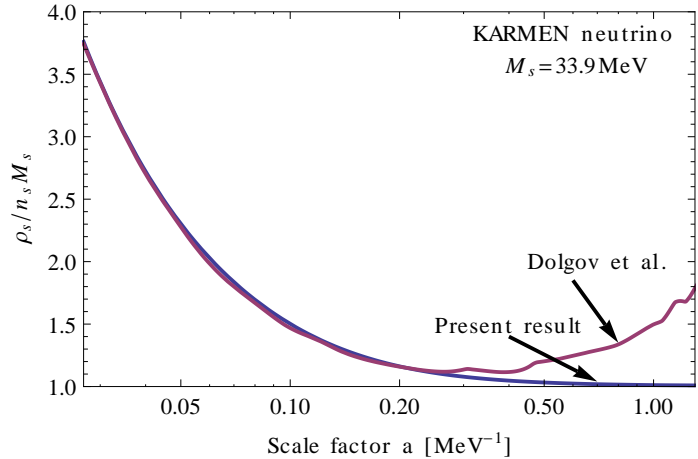}
  \caption{Ratio $\rho_s / n_s M_s$ as a function of scale factor for
    $M_s=33.9\MeV$ sterile neutrino. The upper curve is the result of
    Ref.~\protect\cite{Dolgov:00a}, the lower curve is the present work.}
  \label{fig:rho_o_mn}
\end{figure}

Being in equilibrium the sterile neutrino spectrum becomes more and more
non-relativistic with time due to the redshift. Therefore the ratio $\rho_s /
M_s n_s $ of the energy density $\rho_s$ to the mass times number density
$n_s$ should approach~1 at lower temperatures. We have recomputed the
evolution of the system using our code, without the Boltzmann
approximation. Fig.~\ref{fig:rho_o_mn} shows the comparison of the results
with those of \cite{Dolgov:00a} for sterile neutrino lifetime
$\tau_s=0.3\sec$. Both results coincide till $T \approx 5$ MeV and after that
moment ratio $\rho_s / M_s n_s $ of~\cite{Dolgov:00a} stops decreasing, while
the numerical result we obtained shows the expected behaviour --- the ratio continues to
decrease, approaching~1. 

\subsection{Massive $\nu_\tau$}

\begin{figure}[tp]
  \includegraphics[width=0.5\textwidth]{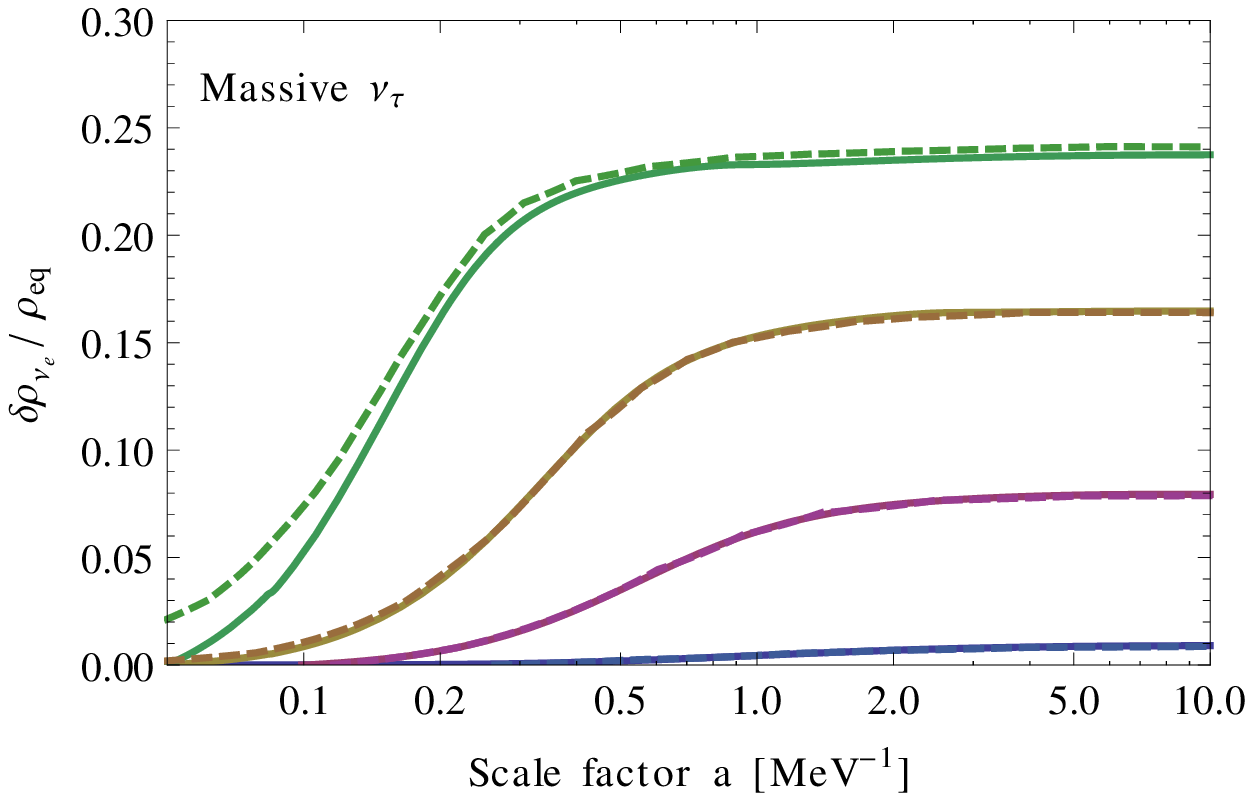}
\includegraphics[width=0.5\textwidth]{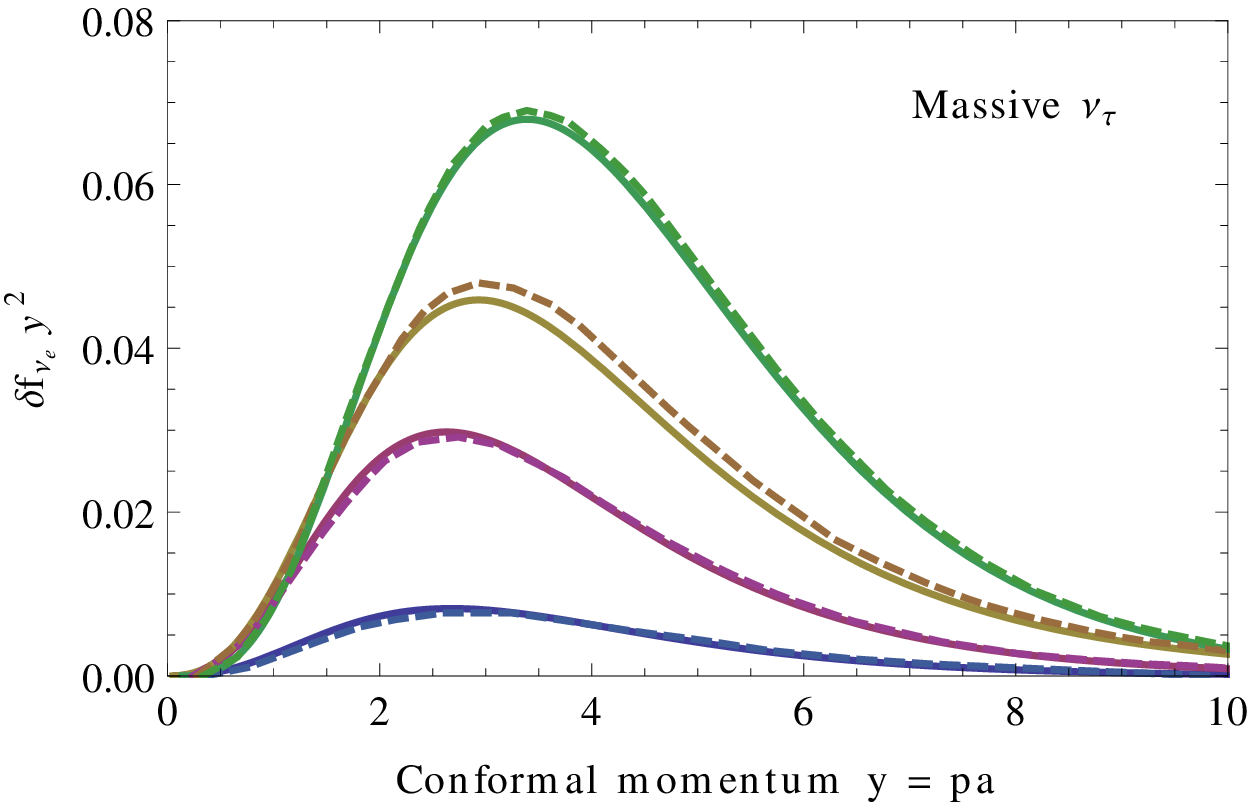}
\caption{\textbf{Left:} Relative deviation from its equilibrium value of
  $\nu_e$ energy density $\delta \rho_{\nu_e} / \rho_{\rm eq}$ in a model
  where tau neutrino is massive. 
\textbf{Right:} Spectrum distortion $y^2 \delta
  f_{\nu_e} / f_{\rm eq}$ for the same model. In both panels $M_{\nu_\tau}=
  0,3,7,20\MeV$ from bottom to top, the solid curves depict the numerical results of this work, and the dashed -- the results of~\protect\cite{Dolgov:1997it}. }
  \label{fig:masstau}
\end{figure}

Next we considered a model with the massive tau
neutrino~\cite{Dolgov:1997it,Kawasaki:1993gz}. Fig.~\ref{fig:masstau} presents relative
deviation of the energy densities of massless neutrinos $\delta \rho_\nu /
\rho_{\rm eq}$ produced by our code and plotted in~\cite{Dolgov:1997it}. $\rho_{\rm
  eq}= \frac{7 \pi^2 T^4 }{120}$ is the equilibrium energy density of one
neutrino specie, and $\delta \rho_\nu =\rho_\nu-\rho^{eq}_{\nu}$. In
Fig.~\ref{fig:masstau} distortion of electron neutrino spectrum $y^2 \delta
f_{\nu_e} / f_{eq}$ is depicted.  Here one observes good agreement between the
results.

\subsection{Late reheating model}

\begin{figure}
\centering
  \includegraphics[width=0.6\textwidth]{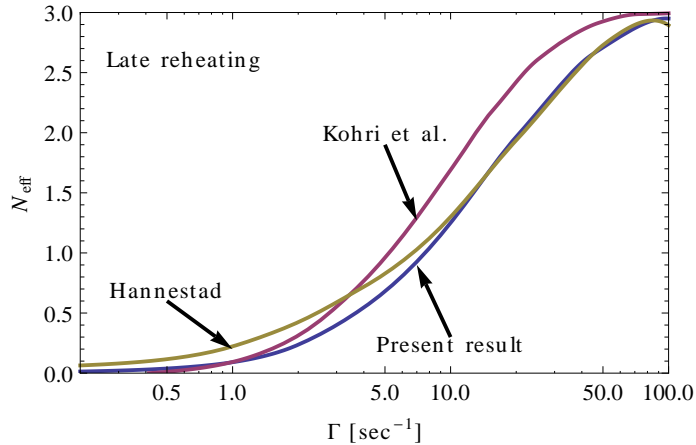}
  \caption{ Effective number of neutrino species $N_{\rm eff}$ depending on
    decay width of heavy non-relativistic particles. Comparison of the results
    of this work and Refs.~\protect\cite{Kawasaki:2000en,Hannestad:04}.}
  \label{fig:Neff_Gamma}
\end{figure}

To test the treatment of MeV decaying particles, we considered the
low-reheating models with the reheating temperature of several MeV
\cite{Kawasaki:2000en,Hannestad:04}. In \cite{Kawasaki:2000en} heavy
non-relativistic particles were considered, that dominated the energy density
of the Universe once and then decayed into electrons, positrons or photons (so
that decay products are quickly thermalized). The most important output is the
effective number of active neutrino species $N_{\rm eff}$ (defined in
Eq.~(\ref{eq:Neff-definition})). Dependence of this quantity on decay width of
heavy particle is presented in Fig.~\ref{fig:Neff_Gamma}. We have noticed some
difference between the results of cited papers and those of our code.  We
believe that this is due to the different approximations made. For example, in
both works \cite{Kawasaki:2000en,Hannestad:04} the scattering processes
involving only neutrinos were not taken into account, approximation of
Boltzmann statistics was used throughout and electron mass was neglected. We
checked that the account of finite electron mass gives a gain of $5\%$ to the
$N_{\rm eff}$ for $\tau=0.1$s, while the account of scatterings involving only
neutrinos gives rise of $1\%$.

\subsection{Instant thermalization of decay products }
\label{sec:decays-thermalization}
Next we considered a model with two heavy Majorana sterile neutrinos, similar
to the \numsm. However, we assumed that for any mass of sterile neutrino it
can decay \textit{only} via channels listed in Table \ref{tab:sterile-decays}
of Appendix \ref{app:matrix-elements}. It is not a natural assumption, because
usually sterile neutrinos heavier than pion decay dominantly into states
containing mesons \cite{Gorbunov:07a}. Also we approximated sterile neutrino
spectrum as a non-relativistic one, while all the other particles are
relativistic and in equilibrium all the time. In this case the system may be
adequately described by the kinetic equation
\begin{equation}
  \label{eq:9}
\frac{d\rho_s}{dt} + 3\frac{\dot a}{a} \rho_s = - \Gamma_s \rho_s
\end{equation}
together with the Friedmann equations
(\ref{eq:Friedmann-first}--\ref{eq:Friedmann-second}). The latter of these
equations can be rewritten as 
\begin{equation}
\label{eq:temperature-decay}
 \frac{d (aT)}{dt} = \frac{30 a\Gamma_s \rho_s}{43 \pi^2 T^3}
\end{equation}
$\Gamma_s$ is the decay width of sterile neutrino, $\rho_s$ is the energy density of sterile neutrinos, and we have used expression for the energy and pressure densities of relativistic species $\rho_{\rm rel}= 3 p_{\rm rel} = 43\pi^2 T^4/120$.

\begin{figure}
  \includegraphics[width=0.5\textwidth]{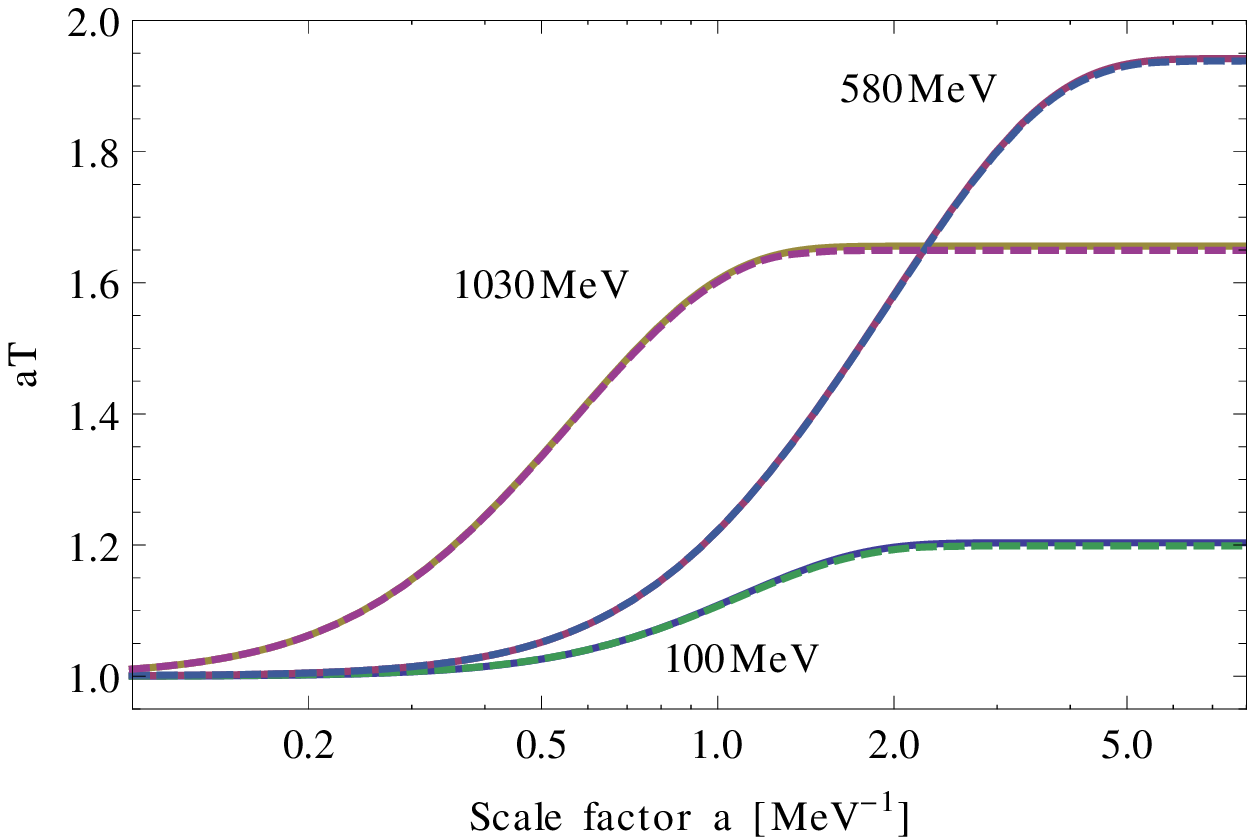}
\includegraphics[width=0.5\textwidth]{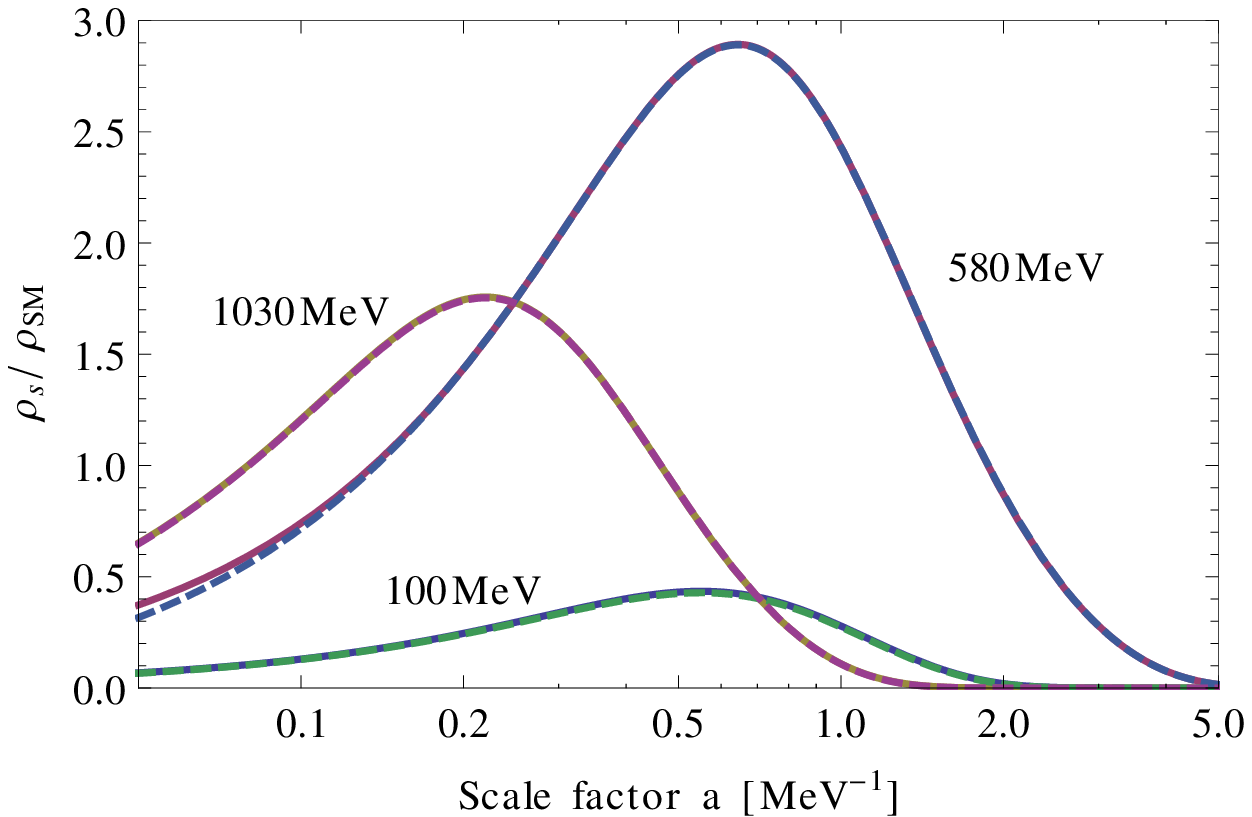}
  \caption{\textbf{Left:} Evolution of $aT$ for the model of Sec.~\protect\ref{sec:decays-thermalization}. \textbf{Right:} $\rho_s / \rho_{SM}$.
We consider three parameter sets: sterile neutrino mass $M_s=580$ MeV with lifetime $\tau=1$sec; $M_s=1030$ MeV with $\tau=0.1$sec; $M_s=100$MeV, $\tau=0.5$sec. The solid line depicts the numerical result of this work, dashed -- the semianalytical calculation.}
  \label{fig:entr_Tx}
\end{figure}

In Figs. \ref{fig:entr_Tx} the evolution of quantities $aT$ and $\rho_s /
\rho_{\rm rel}$ is compared between the results of our code and the
semi-analytic integration of Eqs.~(\ref{eq:9})--(\ref{eq:temperature-decay})
for three different sets of masses and lifetimes. One can see very good
agreement between these results, maximum relative deviation is $1\%$.

\section{Tree-level matrix elements}
\label{app:matrix-elements}

In this Appendix we summarize the matrix elements we used for computing the
collision integrals in Boltzmann equation.  The squares of the matrix elements
for Standard Model particles only are listed in Table
\ref{tab:active-scatterings}, while the squares of the matrix elements of
processes with sterile neutrinos are summarized in
Tables~\ref{tab:sterile-scatterings},~\ref{tab:sterile-decays}. In these
expressions, averaging over helicities of incoming particles and summation
over those of outgoing products is assumed. The reactions are considered for
two cases. In the first one sterile neutrino is a right-chiral Majorana
neutrino that has 2 helicity degrees of freedom. That is actually the case in
our problem, where we have two neutrinos of this kind. The other case
corresponds to sterile neutrino of Dirac nature. Dirac fermions have both
right- and left-chiral components, hence yielding 4 degrees of freedom in
total. Expressions listed in
Tables~\ref{tab:sterile-scatterings},~\ref{tab:sterile-decays} are applicable
for both cases of the neutrino nature. Moreover, to complete the list of
possible tree-level reactions, one has to consider charge-conjugated channels
and take into account that Dirac particle is distinct from its antiparticle,
while Majorana neutrino is not.

 Throughout this Section we use the notations $g_R=\rm{sin}^2 \theta_W$ , $g_L=1/2+\rm{sin}^2 \theta_W$,
$\tilde{g}_L=-1/2+\rm{sin}^2 \theta_W$, where $\theta_W$ is the Weinberg angle
so that $\sin^2 \theta_W \approx 0.23$. The resulting expressions coincide with \cite{Dolgov:97sm,Dolgov:00a}.

 \renewcommand{\arraystretch}{1.2}%

 \begin{table}[t]
 \begin{center}
   \begin{tabular}{|cccc|c|l|}
   \hline
   &  Process ($1+2\to 3+4$) & & & $S$ & $S G_F^{-2} \left|\mathcal{M} \right|^2$  \\
   \hline \hline
   & $\nu_\alpha+\nu_\beta\rightarrow \nu_\alpha + \nu_\beta$ & & & 1 & $32 (p_1 \cdot p_2) (p_3 \cdot p_4)$ \\
   & $\nu_\alpha+\bar{\nu}_\beta \rightarrow \nu_\alpha + \bar{\nu}_\beta$ &  & & 1 & $32 (p_1 \cdot p_4) (p_2 \cdot p_3)$ \\
   & $\nu_\alpha+\nu_\alpha \rightarrow \nu_\alpha + \nu_\alpha$ &  & & 1/2 & $64 (p_1 \cdot p_2) (p_3 \cdot p_4)$ \\
   & $\nu_\alpha+\bar{\nu}_\alpha \rightarrow \nu_\alpha + \bar{\nu}_\alpha$ &  & & 1 & $128 (p_1 \cdot p_4) (p_2 \cdot p_3)$ \\
   & $\nu_\alpha+\bar{\nu}_\alpha \rightarrow \nu_\beta + \bar{\nu}_\beta$ &  & & 1 & $32 (p_1 \cdot p_4) (p_2 \cdot p_3)$ \\
   & $\nu_e+\bar{\nu}_e \rightarrow e^+ + e^-$ &  & & 1 & $128 [ g_L^2 (p_1 \cdot p_4) (p_2 \cdot p_3) +$ \\ 
   &&&&& $g_R^2 (p_1 \cdot p_3)(p_2 \cdot p_4) + g_L g_R m_e^2 (p_1 \cdot p_2) ]$ \\
   & $\nu_e+e^- \rightarrow \nu_e + e^-$ &  & & 1 & $128 [g_L^2 (p_1 \cdot p_2) (p_3 \cdot p_4) +$ \\
   &&&&& $g_R^2 (p_1 \cdot p_4) (p_2 \cdot p_3) - g_L g_R m_e^2 (p_1 \cdot p_3) ]$\\
   & $\nu_e+e^+ \rightarrow \nu_e + e^+$ &  & & 1 & $128 [g_L^2(p_1 \cdot p_4) (p_2 \cdot p_3) +$ \\
   &&&&& $g_R^2 (p_1 \cdot p_2)(p_3 \cdot p_4) - g_L g_R m_e^2 (p_1 \cdot p_3) ]$\\
   & $\nu_{\mu(\tau)}+\bar{\nu}_{\mu(\tau)} \rightarrow e^+ + e^-$ &  & & 1 & $128 [ \tilde{g}_L^2 (p_1 \cdot p_4) (p_2 \cdot p_3) +$ \\ 
   &&&&& $g_R^2 (p_1 \cdot p_3)(p_2 \cdot p_4) + \tilde{g}_L g_R m_e^2 (p_1 \cdot p_2) ]$ \\
   & $\nu_{\mu(\tau)}+e^- \rightarrow \nu_{\mu(\tau)} + e^-$ &  & & 1 & $128 [\tilde{g}_L^2 (p_1 \cdot p_2) (p_3 \cdot p_4) +$ \\
   &&&&& $g_R^2 (p_1 \cdot p_4) (p_2 \cdot p_3) - \tilde{g}_L g_R m_e^2 (p_1 \cdot p_3) ]$\\
   & $\nu_{\mu(\tau)}+e^+ \rightarrow \nu_{\mu(\tau)} + e^+$ &  & & 1 & $128 [\tilde{g}_L^2 (p_1 \cdot p_4) (p_2 \cdot p_3) +$ \\
   &&&&& $g_R^2 (p_1 \cdot p_2)(p_3 \cdot p_4) - \tilde{g}_L g_R m_e^2 (p_1 \cdot p_3) ]$\\
   \hline
 \end{tabular}
\end{center}
\caption{Squared matrix elements for weak processes involving active species
  only. $S$ is the symmetrization factor; $\alpha, \beta = e, \mu, \tau$. In
  all processes we take $\alpha \neq \beta$. The results coincide with those
  of Ref.~\protect\cite{Dolgov:97sm}.}
\label{tab:active-scatterings}
\end{table}

\begin{table}[!thbp]
 \begin{center}
 \begin{tabular}{|cccc|c|l|}
\hline
 &  Process ($1+2\to 3+4$) & & & $S$ & $S G_F^{-2} \left|\mathcal{M} \right|^2$ \\
\hline \hline
  & $\nu_s+\nu_\beta\rightarrow \nu_\alpha + \nu_\beta$ & & & 1 & $32 \vartheta_\alpha^2 (p_1 \cdot p_2) (p_3 \cdot p_4)$ \\
  & $\nu_s+\bar{\nu}_\beta \rightarrow \nu_\alpha + \bar{\nu}_\beta$ &  & & 1 & $32 \vartheta_\alpha^2 (p_1 \cdot p_4) (p_2 \cdot p_3)$ \\
  & $\nu_s+\nu_\alpha \rightarrow \nu_\alpha + \nu_\alpha$ &  & & 1/2 & $64 \vartheta_\alpha^2 (p_1 \cdot p_2) (p_3 \cdot p_4)$ \\
  & $\nu_s+\bar{\nu}_\alpha \rightarrow \nu_\alpha + \bar{\nu}_\alpha$ &  & & 1 & $128\vartheta_\alpha^2 (p_1 \cdot p_4) (p_2 \cdot p_3)$ \\
  & $\nu_s+\bar{\nu}_\alpha \rightarrow \nu_\beta + \bar{\nu}_\beta$ &  & & 1 & $32\vartheta_\alpha^2 (p_1 \cdot p_4) (p_2 \cdot p_3)$ \\
  & $\nu_s+\bar{\nu}_e \rightarrow e^+ + e^-$ &  & & 1 & $128\vartheta_e^2 [ g_L^2 (p_1 \cdot p_4) (p_2 \cdot p_3) +$ \\ 
  &&&&& $g_R^2 (p_1 \cdot p_3)(p_2 \cdot p_4) + g_L g_R m_e^2 (p_1 \cdot p_2) ]$ \\
  & $\nu_s+e^- \rightarrow \nu_e + e^-$ &  & & 1 & $128\vartheta_e^2 [g_L^2 (p_1 \cdot p_2) (p_3 \cdot p_4) +$ \\
  &&&&& $g_R^2 (p_1 \cdot p_4) (p_2 \cdot p_3) - g_L g_R m_e^2 (p_1 \cdot p_3) ]$\\
  & $\nu_s+e^+ \rightarrow \nu_e + e^+$ &  & & 1 & $128\vartheta_e^2 [g_L^2(p_1 \cdot p_4) (p_2 \cdot p_3) +$ \\
  &&&&& $g_R^2 (p_1 \cdot p_2)(p_3 \cdot p_4) - g_L g_R m_e^2 (p_1 \cdot p_3) ]$\\
  & $\nu_s+\bar{\nu}_{\mu(\tau)} \rightarrow e^+ + e^-$ &  & & 1 & $128\vartheta_{\mu(\tau)}^2 [ \tilde{g}_L^2 (p_1 \cdot p_4) (p_2 \cdot p_3) +$ \\ 
  &&&&& $g_R^2 (p_1 \cdot p_3)(p_2 \cdot p_4) + \tilde{g}_L g_R m_e^2 (p_1 \cdot p_2) ]$ \\
  & $\nu_s+e^- \rightarrow \nu_{\mu(\tau)} + e^-$ &  & & 1 & $128\vartheta_{\mu(\tau)}^2  [\tilde{g}_L^2 (p_1 \cdot p_2) (p_3 \cdot p_4) +$ \\
  &&&&& $g_R^2 (p_1 \cdot p_4) (p_2 \cdot p_3) - \tilde{g}_L g_R m_e^2 (p_1 \cdot p_3) ]$\\
  & $\nu_s+e^+ \rightarrow \nu_{\mu(\tau)} + e^+$ &  & & 1 & $128\vartheta_{\mu(\tau)}^2  [\tilde{g}_L^2 (p_1 \cdot p_4) (p_2 \cdot p_3) +$ \\
  &&&&& $g_R^2 (p_1 \cdot p_2)(p_3 \cdot p_4) - \tilde{g}_L g_R m_e^2 (p_1 \cdot p_3) ]$\\
\hline
 \end{tabular}
\end{center}
\caption{Squared matrix elements for \textit{scatterings} of sterile neutrinos
  $\nu_S$. Here $S$ is the symmetrization factor; $\alpha, \beta = e, \mu, \tau$;~
  $\alpha \neq \beta$. $\vartheta_{\alpha}$ is the mixing angle of
  sterile neutrino with $\nu_{\alpha}$. The results are applicable for one right-chiral Majorana neutrino as well as for one Dirac neutrino, for details see text.}
\label{tab:sterile-scatterings}
\end{table}

\begin{table}
 \begin{center}
   \begin{tabular}{|cccc|c|l|}
   \hline
   &  Process ($1\to 2+3+4$) & & & $S$ & $S G_F^{-2} \left|\mathcal{M} \right|^2$ \\
   \hline \hline
   & $ \nu_S \rightarrow  \nu_\alpha + \nu_\beta + \bar{\nu}_\beta $ & & & 1 & $ 32~\vartheta_\alpha^2 (p_1 \cdot p_4)(p_2 \cdot p_3) $ \\
   & $ \nu_S \rightarrow \nu_\alpha + \nu_\alpha + \bar{\nu}_\alpha $ & & & 1/2 & $64~ \vartheta_\alpha^2 (p_1 \cdot p_4)(p_2 \cdot p_3) $ \\
   & $ \nu_S \rightarrow \nu_e + e^{+} + e^{-} $ & & & 1 & $128~ \vartheta_e^2 [g^2_L (p_1 \cdot p_3)(p_2 \cdot p_4) + $ \\
   &&&&& $g^2_R (p_1 \cdot p_4)(p_2 \cdot p_3) + g_L g_R m_e^2 (p_1 \cdot p_2)]$\\
   & $ \nu_S \rightarrow \nu_{\mu (\tau)} + e^{+} + e^{-} $ & & & 1 & $128~ \vartheta_{\mu (\tau)}^2 [\tilde{g}^2_L (p_1 \cdot p_3)(p_2 \cdot p_4) + $ \\
   &&&&& $g^2_R (p_1 \cdot p_4)(p_2 \cdot p_3) + \tilde{g}_L g_R m_e^2 (p_1 \cdot p_2)]$\\
\hline
 \end{tabular}
\end{center}
\caption{
  Squared matrix elements for \textit{decays} of sterile neutrinos
  $\nu_S$. Here $S$ is the symmetrization factor; $\alpha, \beta = e, \mu, \tau$;~
  $\alpha \neq \beta$. $\vartheta_{\alpha}$ is the mixing angle of
  sterile neutrino with $\nu_{\alpha}$. The results are both for Majorana and Dirac neutrinos, for details see text.}
\label{tab:sterile-decays}
\end{table}

\section{Neutrino oscillations}
\label{app:neutrino-oscillations}

The active neutrinos of different flavours $\nu_e,\nu_\mu,\nu_\tau$ are
related to the mass eigen-state basis $\nu_1,\nu_2,\nu_3$ via a non-diagonal
Pontecorvo-Maki-Nakagava-Sakata (PMNS) matrix $V$ $|\nu_\alpha \rangle= \sum
V_{\alpha i} |\nu_i \rangle$ (see e.g.~\cite{Strumia:2006db} for reviews):
\begin{equation}
\nonumber
V= \left(\begin{array}{ccc} 1&0&0\\ 0&c_{23}&s_{23}\\ 0&-s_{23}&c_{23} \end{array}\right)
\left(\begin{array}{ccc} c_{13}&0&s_{13}\\ 0&e^{i\phi}&0 \\ -s_{13}&0&c_{13} \end{array}\right)
\left(\begin{array}{ccc} c_{12}&s_{12}&0\\ -s_{12}&c_{12}&0 \\ 0&0&1 \end{array}\right)\;.
\end{equation}
here $c_{ij}=\cos \theta_{ij}$ and $s_{ij}=\sin \theta_{ij}$ are functions of
the active-active neutrino mixing angles $\theta_{ij}$.

Exact treatment of active neutrino oscillation in the early Universe is a
difficult task~(see e.g.~\cite{Dolgov:2002ab,Dolgov:04,Kirilova:09}) Characteristic
timescale of oscillation between $i$ and $j$ mass eigen-states for a neutrino
with energy $E$ is \cite{Strumia:2006db}
\begin{equation}
\label{eq:oscillation-timescale}
\tau_{ij} = \frac{4\pi E}{|m_i^2-m_j^2|} \approx 8.3\times 10^{-6}{\rm s}~\frac{E}{\rm MeV}~\frac{10^{-3} \eV^2}{|m_i^2-m_j^2|}
\end{equation} 
Average energy of relativistic Fermi particles in equilibrium is $\langle E \rangle = 3.15 T$ \cite{Kolb:90}. Applying this relation to active neutrinos and using their measured mass differences \cite{Schwetz:2011zk} $m_2^2-m_1^2 \approx 7.6\times 10^{-5}\eV^2,~|m_3^2-m_1^2|\approx 2.5\times 10^{-3}\eV^2$, we obtain
\begin{equation}
  \tau_{12}\approx 1.0\times 10^{-3}\sec\frac{T}{3\MeV}\;,~~\tau_{13}\approx 3.1\times 10^{-5}\sec\frac{T}{3\MeV}\;,
\end{equation} 
provided that influence of the surrounding environment on neutrino propagation is neglected.  One sees therefore that about the moment active neutrino decouples $T\simeq 3\MeV$ typical oscillation timescales are much smaller than the Hubble expansion time given by Eq.~(\ref{eq:Friedmann-first})
\begin{equation}
\tau_H= \sqrt{\frac{15}{4\pi^3 g_* G_N T^4}} \simeq 0.16\sec \left( \frac{3\MeV}{T}  \right)^2\;.
\end{equation}
Here $g_*\approx 11$ (at $T\sim \MeV$) \cite{Kolb:90} is the so-called number of relativistic species that enters energy-temperature relation $\rho = \frac{\pi^2 g_* T^4}{30}$. Therefore, active neutrinos oscillate many times between the subsequent reactions involving them. In quantitative terms it means that probabilities $P_{\alpha\beta}$ to transform from flavour $\alpha$ to flavour $\beta$ are oscillating functions of time. In realistic situation neutrinos do not have a definite momentum but are created in wave packets that are superpositions of states which have one. Since oscillation periods are momentum-dependent according to Eq.~(\ref{eq:oscillation-timescale}), each state in the superposition will have his own period. Therefore after sufficiently many periods initial phases characterizing superposition will change, and there is no reason for the phase changes to be correlated with each other. So the decoherence of states is what happens. This phenomenon can be described effectively by averaging transition probabilities $P_{\alpha\beta}$ over time. Resulting expressions are~\cite{Strumia:2006db}
\begin{subequations}
\label{eq:oscillation-probabilities}
\begin{align}
P_{ee} &= 1 - \frac{1}{2}(\sin^2 2\theta_{13} + \cos^4 \theta_{13} \sin^2
2\theta_{12})\\
P_{e\mu} &= P_{\mu e} = \frac{1}{2} \cos^2 \theta_{13} \sin^2 2\theta_{12}\\
P_{e\tau} &= P_{\tau e} = \sin^2 \theta_{13} \cos^2 \theta_{13} \left(2 -
\frac{1}{2} \sin^2 2\theta_{12}\right) \\
P_{\mu\mu} &= 1 - \frac{1}{2}\sin^2 2\theta_{12}\\
P_{\mu\tau} &= P_{\tau\mu} = \frac{1}{2}\sin^2 \theta_{13} \sin^2 2
\theta_{12}\\
P_{\tau\tau} &= 1 - \sin^2 \theta_{13} \left( 2\cos^2 \theta_{13} + \frac{1}{2}
\sin^2 \theta_{13} \sin^2 2\theta_{12} \right)
\end{align}
\end{subequations}

To understand what happens with a neutrino, consider example of electron-neutrino created in electron-positron annihilation. At the production time this particle has probability 1 to oscillate into $\nu_e$ and zero for other final state. After long enough time for many oscillations to happen and before the time when a collision with other particle becomes quite probable, the decoherence comes into play. So now we may find the $\nu_e$ with probability $P_{ee}$, $\nu_\mu$ with probability $P_{e\mu}$ and $\nu_\tau$ with $P_{e \tau}$. The production rate of the initial specimen per unit time is proportional to collision integral $I_e$, according to the Boltzmann equation (\ref{eq:Boltzmann-equation}). But the actual number of produced electron neutrinos 
is actually reduced by factor $P_{ee}$. And even if (consider this hypothetical situation) muon neutrino does not interact with plasma, it will be anyway produced, at rate $P_{e\mu} I_e$. Generalization to other neutrino flavours leads us to conclusion that the modified Boltzmann equation
\begin{equation}
  \frac{df_\alpha}{dt} = \sum P_{\alpha\beta} I_\beta
\end{equation}
describes neutrino dynamics correctly (that is not the case for the initial
equation (\ref{eq:Boltzmann-equation})). For the actual computations we use the following experimental best-fit values: $\sin^2\theta_{12} = 0.31,~\sin^2\theta_{23}=0.52$ from~\cite{Schwetz:2011zk}, and $\sin^2 2\theta_{13} = 0.09$ from the Daya Bay~\cite{An:2012eh}. The latter number is close to the result $\sin^22\theta_{13}=0.11$ indicated by another recent experiment, RENO \cite{Ahn:2012nd}.

However, in dense medium oscillations proceed differently due to considerable
effects of the plasma on properties of a single particle. Still, the
phenomenon can be described by the formalism of the PMNS matrix. The
difference is that mixing parameters together with masses now depend on
properties of the environment. In case of plasma close to equilibrium with no
non-trivial conserving charges present the parameter describing it is the
temperature. So the parameters of the PMNS become
\emph{temperature-dependent}. In language of the effective Hamiltonian
approach the system of three neutrinos is described by the addition of medium
potential $\Delta H_M$ to the Hamiltonian $H_{\rm V}$ of the system in vacuum
\cite{Notzold:87}

\begin{equation}
H_M = H_{\rm V} + \Delta H_M,~~H_{\rm V}=\frac{1}{2E} V^* \diag(m_1^2,m_2^2,m_3^2) V^\dagger \;,
\end{equation}
where $E$ is the neutrino energy. Diagonalization of the total propagation Hamiltonian $H_M$ gives effective masses and mixings.

The medium potential comprises effects of neutrino interactions. Since neutrinos take part only in charged- and neutral-current interactions, matter potential has two terms $\Delta H_{CC}$ and $\Delta H_{NC}$, respectively.  All neutrinos couple to neutral currents identically, so $\Delta H_{NC}$ is proportional to unit matrix. Therefore this term just renormalizes energy, and does not affect oscillations. In contrast, the charged-current term is non-diagonal and is present only for $\nu_e$. The reason is that due to abundance of electrons in plasma, $\nu_e$ couples effectively to charged currents, while at temperatures below the muon's mass there is no significant contribution of muons and tau-leptons to realize coupling of other neutrinos to W boson.

 Explicitly matter potential is \cite{Strumia:2006db}
\begin{equation}
\Delta H_{CC} = -\frac{14\sqrt{2} G_F}{45 M_W^2}~ E~T^4 ~\diag(1,0,0)
\end{equation}
in the flavour neutrino basis ($\nu_e,\nu_\mu,\nu_\tau$). $M_W$ is the mass of the W-boson.

So far we have dropped sterile neutrinos from consideration. But their mixing
properties are also altered in hot plasma. Using the approach of the effective
Hamiltonian for them, one finds that their effective mixing angles in medium
$\theta_M$ differ from that in vacuum $\theta_V$ as \cite{Notzold:87}
\begin{equation}
\frac{\theta_M - \theta_V}{\theta_V} \sim \frac{G_F T^5}{M_W^2 M_S^2} \sim 10^{-11} \times \left(\frac{T}{100\MeV}\right)^6~ \left( \frac{10\MeV}{M_S} \right)^2\
\end{equation}
for small mixing angles $\theta_V$. Therefore the mixing angle is not altered
significantly for sterile neutrinos and matter effects are negligible for
their dynamics.

\let\jnlstyle=\rm\def\jref#1{{\jnlstyle#1}}\def\aj{\jref{AJ}}
  \def\araa{\jref{ARA\&A}} \def\apj{\jref{ApJ}\ } \def\apjl{\jref{ApJ}\ }
  \def\apjs{\jref{ApJS}} \def\ao{\jref{Appl.~Opt.}} \def\apss{\jref{Ap\&SS}}
  \def\aap{\jref{A\&A}} \def\aapr{\jref{A\&A~Rev.}} \def\aaps{\jref{A\&AS}}
  \def\azh{\jref{AZh}} \def\baas{\jref{BAAS}} \def\jrasc{\jref{JRASC}}
  \def\memras{\jref{MmRAS}} \def\mnras{\jref{MNRAS}\ }
  \def\pra{\jref{Phys.~Rev.~A}\ } \def\prb{\jref{Phys.~Rev.~B}\ }
  \def\prc{\jref{Phys.~Rev.~C}\ } \def\prd{\jref{Phys.~Rev.~D}\ }
  \def\pre{\jref{Phys.~Rev.~E}} \def\prl{\jref{Phys.~Rev.~Lett.}}
  \def\pasp{\jref{PASP}} \def\pasj{\jref{PASJ}} \def\qjras{\jref{QJRAS}}
  \def\skytel{\jref{S\&T}} \def\solphys{\jref{Sol.~Phys.}}
  \def\sovast{\jref{Soviet~Ast.}} \def\ssr{\jref{Space~Sci.~Rev.}}
  \def\zap{\jref{ZAp}} \def\nat{\jref{Nature}\ } \def\iaucirc{\jref{IAU~Circ.}}
  \def\aplett{\jref{Astrophys.~Lett.}}
  \def\apspr{\jref{Astrophys.~Space~Phys.~Res.}}
  \def\bain{\jref{Bull.~Astron.~Inst.~Netherlands}}
  \def\fcp{\jref{Fund.~Cosmic~Phys.}} \def\gca{\jref{Geochim.~Cosmochim.~Acta}}
  \def\grl{\jref{Geophys.~Res.~Lett.}} \def\jcp{\jref{J.~Chem.~Phys.}}
  \def\jgr{\jref{J.~Geophys.~Res.}}
  \def\jqsrt{\jref{J.~Quant.~Spec.~Radiat.~Transf.}}
  \def\memsai{\jref{Mem.~Soc.~Astron.~Italiana}}
  \def\nphysa{\jref{Nucl.~Phys.~A}} \def\physrep{\jref{Phys.~Rep.}}
  \def\physscr{\jref{Phys.~Scr}} \def\planss{\jref{Planet.~Space~Sci.}}
  \def\procspie{\jref{Proc.~SPIE}} \let\astap=\aap \let\apjlett=\apjl
  \let\apjsupp=\apjs \let\applopt=\ao
\providecommand{\href}[2]{#2}\begingroup\raggedright\endgroup


\begin{thebibliography}{10}

\bibitem{Izotov:2010ca}
Y.~Izotov and T.~Thuan, {\it {The primordial abundance of 4He: evidence for
  non-standard big bang nucleosynthesis}},  {\em Astrophys.J.} {\bf 710} (2010)
  L67--L71 [\href{http://arXiv.org/abs/1001.4440}{{\tt 1001.4440}}].

\bibitem{WMAP7}
E.~{Komatsu}, K.~M. {Smith}, J.~{Dunkley}, C.~L. {Bennett}, B.~{Gold},
  G.~{Hinshaw}, N.~{Jarosik}, D.~{Larson}, M.~R. {Nolta}, L.~{Page}, D.~N.
  {Spergel}, M.~{Halpern}, R.~S. {Hill}, A.~{Kogut}, M.~{Limon}, S.~S. {Meyer},
  N.~{Odegard}, G.~S. {Tucker}, J.~L. {Weiland}, E.~{Wollack} and E.~L.
  {Wright}, {\it {Seven-year Wilkinson Microwave Anisotropy Probe (WMAP)
  Observations: Cosmological Interpretation}},  {\em \apjs} {\bf 192} (Feb.,
  2011) 18--+ [\href{http://arXiv.org/abs/1001.4538}{{\tt 1001.4538}}].

\bibitem{Alpher:1948ve}
R.~Alpher, H.~Bethe and G.~Gamow, {\it {The origin of chemical elements}},
  {\em Phys.Rev.} {\bf 73} (1948) 803--804.

\bibitem{Iocco:08}
F.~Iocco, G.~Mangano, G.~Miele, O.~Pisanti and P.~D. Serpico, {\it {Primordial
  Nucleosynthesis: from precision cosmology to fundamental physics}},  {\em
  Phys. Rept.} {\bf 472} (2009) 1--76
  [\href{http://arXiv.org/abs/0809.0631}{{\tt 0809.0631}}].

\bibitem{Steigman:07}
G.~Steigman, {\it {Primordial Nucleosynthesis in the Precision Cosmology Era}},
   {\em Ann. Rev. Nucl. Part. Sci.} {\bf 57} (2007) 463--491
  [\href{http://arXiv.org/abs/0712.1100}{{\tt 0712.1100}}].

\bibitem{Pospelov:10b}
M.~Pospelov and J.~Pradler, {\it {Big Bang Nucleosynthesis as a Probe of New
  Physics}},  {\em Ann. Rev. Nucl. Part. Sci.} {\bf 60} (2010) 539--568
  [\href{http://arXiv.org/abs/1011.1054}{{\tt 1011.1054}}].

\bibitem{Boyarsky:09a}
A.~Boyarsky, O.~Ruchayskiy and M.~Shaposhnikov, {\it {The role of sterile
  neutrinos in cosmology and astrophysics}},  {\em Ann. Rev. Nucl. Part. Sci.}
  {\bf 59} (2009) 191 [\href{http://arXiv.org/abs/0901.0011}{{\tt 0901.0011}}].

\bibitem{Kusenko:09a}
A.~Kusenko, {\it {Sterile neutrinos: the dark side of the light fermions}},
  {\em Phys. Rept.} {\bf 481} (2009) 1--28
  [\href{http://arXiv.org/abs/0906.2968}{{\tt 0906.2968}}].

\bibitem{Akhmedov:98}
E.~K. Akhmedov, V.~A. Rubakov and A.~Y. Smirnov, {\it Baryogenesis via neutrino
  oscillations},  {\em Phys. Rev. Lett.} {\bf 81} (1998) 1359--1362
  [\href{http://arXiv.org/abs/hep-ph/9803255}{{\tt hep-ph/9803255}}].

\bibitem{Asaka:05a}
T.~Asaka, S.~Blanchet and M.~Shaposhnikov, {\it The {nuMSM}, dark matter and
  neutrino masses},  {\em Phys. Lett.} {\bf B631} (2005) 151--156
  [\href{http://arXiv.org/abs/hep-ph/0503065}{{\tt hep-ph/0503065}}].

\bibitem{Asaka:05b}
T.~{Asaka} and M.~{Shaposhnikov}, {\it {The {nuMSM}, dark matter and baryon
  asymmetry of the universe}},  {\em Phys. Lett. B} {\bf 620} (July, 2005)
  17--26 [\href{http://arXiv.org/abs/arXiv:hep-ph/0505013}{{\tt
  arXiv:hep-ph/0505013}}].

\bibitem{Shaposhnikov:08a}
M.~{Shaposhnikov}, {\it {The nuMSM, leptonic asymmetries, and properties of
  singlet fermions}},  {\em JHEP} {\bf 08} (2008) 008
  [\href{http://arXiv.org/abs/0804.4542}{{\tt 0804.4542}}].

\bibitem{Canetti:10a}
L.~Canetti and M.~Shaposhnikov, {\it {Baryon Asymmetry of the Universe in the
  NuMSM}},  {\em JCAP} {\bf 1009} (2010) 001
  [\href{http://arXiv.org/abs/1006.0133}{{\tt 1006.0133}}].

\bibitem{Laine:08a}
M.~{Laine} and M.~{Shaposhnikov}, {\it {Sterile neutrino dark matter as a
  consequence of {$\nu$}MSM-induced lepton asymmetry}},  {\em JCAP} {\bf 6}
  (June, 2008) 31--+ [\href{http://arXiv.org/abs/arXiv:0804.4543}{{\tt
  arXiv:0804.4543}}].

\bibitem{Dolgov:00a}
A.~D. Dolgov, S.~H. Hansen, G.~Raffelt and D.~V. Semikoz, {\it {Cosmological
  and astrophysical bounds on a heavy sterile neutrino and the KARMEN
  anomaly}},  {\em Nucl. Phys.} {\bf B580} (2000) 331--351
  [\href{http://arXiv.org/abs/hep-ph/0002223}{{\tt hep-ph/0002223}}].

\bibitem{Dolgov:00b}
A.~D. Dolgov, S.~H. Hansen, G.~Raffelt and D.~V. Semikoz, {\it {Heavy sterile
  neutrinos: Bounds from big-bang nucleosynthesis and SN 1987A}},  {\em Nucl.
  Phys.} {\bf B590} (2000) 562--574
  [\href{http://arXiv.org/abs/hep-ph/0008138}{{\tt hep-ph/0008138}}].

\bibitem{Smith:08}
C.~J. Smith, G.~M. Fuller and M.~S. Smith, {\it {Big Bang Nucleosynthesis with
  Independent Neutrino Distribution Functions}},  {\em Phys.Rev.} {\bf D79}
  (2009) 105001 [\href{http://arXiv.org/abs/0812.1253}{{\tt 0812.1253}}].

\bibitem{Fuller:2011qy}
G.~M. Fuller, C.~T. Kishimoto and A.~Kusenko, {\it {Heavy sterile neutrinos,
  entropy and relativistic energy production, and the relic neutrino
  background}},  \href{http://arXiv.org/abs/1110.6479}{{\tt 1110.6479}}.

\bibitem{Dolgov:1997it}
A.~Dolgov, S.~Hansen and D.~Semikoz, {\it {Impact of massive tau neutrinos on
  primordial nucleosynthesis. Exact calculations}},  {\em Nucl.Phys.} {\bf
  B524} (1998) 621--638 [\href{http://arXiv.org/abs/hep-ph/9712284}{{\tt
  hep-ph/9712284}}].

\bibitem{Dolgov:88}
A.~Dolgov and D.~Kirilova, {\it Nonequilibrium decays of light particles and
  the primordial nucleosynthesis},  {\em Int.J.Mod.Phys.} {\bf A3} (1988) 267.

\bibitem{Kawasaki:1993gz}
M.~Kawasaki, P.~Kernan, H.-S. Kang, R.~J. Scherrer, G.~Steigman {\em et.~al.},
  {\it {Big bang nucleosynthesis constraints on the tau-neutrino mass}},  {\em
  Nucl.Phys.} {\bf B419} (1994) 105--128.

\bibitem{Kawasaki:2000en}
M.~Kawasaki, K.~Kohri and N.~Sugiyama, {\it {MeV scale reheating temperature
  and thermalization of neutrino background}},  {\em Phys.Rev.} {\bf D62}
  (2000) 023506 [\href{http://arXiv.org/abs/astro-ph/0002127}{{\tt
  astro-ph/0002127}}].

\bibitem{Hannestad:04}
S.~Hannestad, {\it {What is the lowest possible reheating temperature?}},  {\em
  Phys. Rev.} {\bf D70} (2004) 043506
  [\href{http://arXiv.org/abs/astro-ph/0403291}{{\tt astro-ph/0403291}}].

\bibitem{Dolgov:98addendum}
A.~D. Dolgov, S.~H. Hansen and D.~V. Semikoz, {\it {Nonequilibrium corrections
  to the spectra of massless neutrinos in the early universe. (Addendum)}},
  {\em Nucl. Phys.} {\bf B543} (1999) 269--274
  [\href{http://arXiv.org/abs/hep-ph/9805467}{{\tt hep-ph/9805467}}].

\bibitem{Dolgov:97sm}
A.~D. Dolgov, S.~H. Hansen and D.~V. Semikoz, {\it {Non-equilibrium corrections
  to the spectra of massless neutrinos in the early universe}},  {\em Nucl.
  Phys.} {\bf B503} (1997) 426--444
  [\href{http://arXiv.org/abs/hep-ph/9703315}{{\tt hep-ph/9703315}}].

\bibitem{Gorbunov:07a}
D.~Gorbunov and M.~Shaposhnikov, {\it How to find neutral leptons of the
  numsm?},  {\em JHEP} {\bf 10} (2007) 015
  [\href{http://arXiv.org/abs/arXiv:0705.1729 [hep-ph]}{{\tt arXiv:0705.1729
  [hep-ph]}}].

\bibitem{Nakamura:2010zzi}
{\bf Particle Data Group} Collaboration, K.~Nakamura {\em et.~al.}, {\it
  {Review of particle physics}},  {\em J.Phys.G} {\bf G37} (2010) 075021.

\bibitem{Serebrov:2010sg}
A.~Serebrov and A.~Fomin, {\it {Neutron lifetime from a new evaluation of
  ultracold neutron storage experiments}},  {\em Phys.Rev.} {\bf C82} (2010)
  035501 [\href{http://arXiv.org/abs/1005.4312}{{\tt 1005.4312}}].

\bibitem{Lesgourgues:99}
J.~{Lesgourgues} and S.~{Pastor}, {\it {Cosmological implications of a relic
  neutrino asymmetry}},  {\em \prd} {\bf 60} (Nov., 1999) 103521--+
  [\href{http://arXiv.org/abs/hep-ph/9904411}{{\tt hep-ph/9904411}}].

\bibitem{Serpico:05}
P.~D. Serpico and G.~G. Raffelt, {\it {Lepton asymmetry and primordial
  nucleosynthesis in the era of precision cosmology}},  {\em Phys. Rev.} {\bf
  D71} (2005) 127301 [\href{http://arXiv.org/abs/astro-ph/0506162}{{\tt
  astro-ph/0506162}}].

\bibitem{Smith:2006uw}
C.~J. Smith, G.~M. Fuller, C.~T. Kishimoto and K.~N. Abazajian, {\it {Light
  Element Signatures of Sterile Neutrinos and Cosmological Lepton Numbers}},
  {\em Phys.Rev.} {\bf D74} (2006) 085008
  [\href{http://arXiv.org/abs/astro-ph/0608377}{{\tt astro-ph/0608377}}].

\bibitem{Mangano:10}
G.~Mangano, G.~Miele, S.~Pastor, O.~Pisanti and S.~Sarikas, {\it {Constraining
  the cosmic radiation density due to lepton number with Big Bang
  Nucleosynthesis}},  {\em JCAP} {\bf 1103} (2011) 035
  [\href{http://arXiv.org/abs/1011.0916}{{\tt 1011.0916}}].

\bibitem{Kolb:90}
E.~Kolb and M.~Turner, {\em The Early Universe}.
\newblock Ad{\-d}i{\-s}on-Wes{\-l}ey, Reading, MA, USA, 1990.
\newblock Prepared with {\LaTeX}.

\bibitem{Bezrukov:08a}
F.~Bezrukov, D.~Gorbunov and M.~Shaposhnikov, {\it {On initial conditions for
  the Hot Big Bang}},  {\em JCAP} {\bf 0906} (2009) 029
  [\href{http://arXiv.org/abs/0812.3622}{{\tt 0812.3622}}].

\bibitem{Kawano:1988vh}
L.~Kawano, {\it {Let's Go: Early Universe. Guide to Primordial Nucleosynthesis
  Programming}}, .

\bibitem{Kawano:1992ua}
L.~Kawano, {\it {Let's go: Early universe. 2. Primordial nucleosynthesis: The
  Computer way}}, .

\bibitem{Kernan:1994je}
P.~J. Kernan and L.~M. Krauss, {\it {Refined big bang nucleosynthesis
  constraints on Omega (baryon) and N (neutrino)}},  {\em Phys.Rev.Lett.} {\bf
  72} (1994) 3309--3312 [\href{http://arXiv.org/abs/astro-ph/9402010}{{\tt
  astro-ph/9402010}}].

\bibitem{Boyd:2010kj}
R.~N. Boyd, C.~R. Brune, G.~M. Fuller and C.~J. Smith, {\it {New Nuclear
  Physics for Big Bang Nucleosynthesis}},  {\em Phys.Rev.} {\bf D82} (2010)
  105005 [\href{http://arXiv.org/abs/1008.0848}{{\tt 1008.0848}}].

\bibitem{Fuller:2010un}
G.~M. Fuller and C.~J. Smith, {\it {Nuclear weak interaction rates in
  primordial nucleosynthesis}},  {\em Phys.Rev.} {\bf D82} (2010) 125017
  [\href{http://arXiv.org/abs/1009.0277}{{\tt 1009.0277}}].

\bibitem{Coc:2011az}
A.~Coc, S.~Goriely, Y.~Xu, M.~Saimpert and E.~Vangioni, {\it {Standard Big-Bang
  Nucleosynthesis up to CNO with an improved extended nuclear network}},  {\em
  Astrophys.J.} {\bf 744} (2012) 158
  [\href{http://arXiv.org/abs/1107.1117}{{\tt 1107.1117}}].

\bibitem{Sarkar:1995dd}
S.~Sarkar, {\it {Big bang nucleosynthesis and physics beyond the standard
  model}},  {\em Rept.Prog.Phys.} {\bf 59} (1996) 1493--1610
  [\href{http://arXiv.org/abs/hep-ph/9602260}{{\tt hep-ph/9602260}}]. Dedicated
  to Dennis Sciama on his 67th birthday.

\bibitem{Pisanti:2007hk}
O.~Pisanti, A.~Cirillo, S.~Esposito, F.~Iocco, G.~Mangano {\em et.~al.}, {\it
  {PArthENoPE: Public Algorithm Evaluating the Nucleosynthesis of Primordial
  Elements}},  {\em Comput.Phys.Commun.} {\bf 178} (2008) 956--971
  [\href{http://arXiv.org/abs/0705.0290}{{\tt 0705.0290}}].

\bibitem{Aver:2011bw}
E.~Aver, K.~A. Olive and E.~D. Skillman, {\it {An MCMC determination of the
  primordial helium abundance}},  \href{http://arXiv.org/abs/1112.3713}{{\tt
  1112.3713}}.

\bibitem{Mangano:11}
G.~Mangano and P.~D. Serpico, {\it {A robust upper limit on $N_{\rm eff}$ from
  BBN, circa 2011}},  {\em Phys.Lett.} {\bf B701} (2011) 296--299
  [\href{http://arXiv.org/abs/1103.1261}{{\tt 1103.1261}}].

\bibitem{Peimbert:2007vm}
M.~Peimbert, V.~Luridiana and A.~Peimbert, {\it {Revised Primordial Helium
  Abundance Based on New Atomic Data}},  {\em Astrophys.J.} {\bf 666} (2007)
  636--646 [\href{http://arXiv.org/abs/astro-ph/0701580}{{\tt
  astro-ph/0701580}}].

\bibitem{Dunkley:2010ge}
J.~Dunkley, R.~Hlozek, J.~Sievers, V.~Acquaviva, P.~Ade {\em et.~al.}, {\it
  {The Atacama Cosmology Telescope: Cosmological Parameters from the 2008 Power
  Spectra}},  {\em Astrophys.J.} {\bf 739} (2011) 52
  [\href{http://arXiv.org/abs/1009.0866}{{\tt 1009.0866}}].

\bibitem{Keisler:2011aw}
R.~Keisler, C.~Reichardt, K.~Aird, B.~Benson, L.~Bleem {\em et.~al.}, {\it {A
  Measurement of the Damping Tail of the Cosmic Microwave Background Power
  Spectrum with the South Pole Telescope}},  {\em Astrophys.J.} {\bf 743}
  (2011) 28 [\href{http://arXiv.org/abs/1105.3182}{{\tt 1105.3182}}].

\bibitem{Gelmini:04}
G.~Gelmini, S.~Palomares-Ruiz and S.~Pascoli, {\it Low reheating temperature
  and the visible sterile neutrino},  {\em Phys. Rev. Lett.} {\bf 93} (2004)
  081302 [\href{http://arXiv.org/abs/astro-ph/0403323}{{\tt
  astro-ph/0403323}}].

\bibitem{Gelmini:2009xd}
G.~B. Gelmini, E.~Osoba and S.~Palomares-Ruiz, {\it {Inert-Sterile Neutrino:
  Cold or Warm Dark Matter Candidate}},
  \href{http://arXiv.org/abs/0912.2478}{{\tt 0912.2478}}.

\bibitem{Fuller:2009zz}
G.~M. Fuller, A.~Kusenko and K.~Petraki, {\it {Heavy sterile neutrinos and
  supernova explosions}},  {\em Phys.Lett.} {\bf B670} (2009) 281--284
  [\href{http://arXiv.org/abs/0806.4273}{{\tt 0806.4273}}].

\bibitem{Benson:2011ut}
B.~Benson, T.~de~Haan, J.~Dudley, C.~Reichardt, K.~Aird {\em et.~al.}, {\it
  {Cosmological Constraints from Sunyaev-Zel'dovich-Selected Clusters with
  X-ray Observations in the First 178 Square Degrees of the South Pole
  Telescope Survey}},  \href{http://arXiv.org/abs/1112.5435}{{\tt 1112.5435}}.

\bibitem{Moresco:2012by}
M.~Moresco, L.~Verde, L.~Pozzetti, R.~Jimenez and A.~Cimatti, {\it {New
  constraints on cosmological parameters and neutrino properties using the
  expansion rate of the Universe to z~1.75}},
  \href{http://arXiv.org/abs/1201.6658}{{\tt 1201.6658}}.

\bibitem{Asaka:06}
T.~Asaka, M.~Shaposhnikov and A.~Kusenko, {\it Opening a new window for warm
  dark matter},  {\em Phys. Lett.} {\bf B638} (2006) 401--406
  [\href{http://arXiv.org/abs/hep-ph/0602150}{{\tt hep-ph/0602150}}].

\bibitem{Asaka:2011pb}
T.~Asaka, S.~Eijima and H.~Ishida, {\it {Mixing of Active and Sterile
  Neutrinos}},  {\em JHEP} {\bf 1104} (2011) 011
  [\href{http://arXiv.org/abs/1101.1382}{{\tt 1101.1382}}].

\bibitem{Ruchayskiy:2011aa}
O.~Ruchayskiy and A.~Ivashko, {\it {Experimental bounds on sterile neutrino
  mixing angles}},  \href{http://arXiv.org/abs/1112.3319}{{\tt 1112.3319}}.

\bibitem{Racle:08}
J.~Racle, {\it Deriving bounds on interactions of numsm sterile neutrinos using
  primordial nucleosynthesis},  Master's thesis, EPFL, 2008.

\bibitem{Mangano:2005cc}
G.~Mangano, G.~Miele, S.~Pastor, T.~Pinto, O.~Pisanti {\em et.~al.}, {\it
  {Relic neutrino decoupling including flavor oscillations}},  {\em Nucl.Phys.}
  {\bf B729} (2005) 221--234 [\href{http://arXiv.org/abs/hep-ph/0506164}{{\tt
  hep-ph/0506164}}].

\bibitem{Strumia:2006db}
A.~Strumia and F.~Vissani, {\it Neutrino masses and mixings and.},
  \href{http://arXiv.org/abs/hep-ph/0606054}{{\tt hep-ph/0606054}}.

\bibitem{Dolgov:2002ab}
A.~Dolgov, S.~Hansen, S.~Pastor, S.~Petcov, G.~Raffelt {\em et.~al.}, {\it
  {Cosmological bounds on neutrino degeneracy improved by flavor
  oscillations}},  {\em Nucl.Phys.} {\bf B632} (2002) 363--382
  [\href{http://arXiv.org/abs/hep-ph/0201287}{{\tt hep-ph/0201287}}].

\bibitem{Dolgov:04}
A.~Dolgov and F.~Villante, {\it {BBN bounds on active sterile neutrino
  mixing}},  {\em Nucl.Phys.} {\bf B679} (2004) 261--298
  [\href{http://arXiv.org/abs/hep-ph/0308083}{{\tt hep-ph/0308083}}].

\bibitem{Kirilova:09}
D.~Kirilova, {\it {Non-equilibrium neutrino in the early universe plasma}},
  {\em AIP Conf.Proc.} {\bf 1121} (2009) 83--89.

\bibitem{Schwetz:2011zk}
T.~Schwetz, M.~Tortola and J.~Valle, {\it {Where we are on $\theta_{13}$:
  addendum to 'Global neutrino data and recent reactor fluxes: status of
  three-flavour oscillation parameters'}},  {\em New J.Phys.} {\bf 13} (2011)
  109401 [\href{http://arXiv.org/abs/1108.1376}{{\tt 1108.1376}}].

\bibitem{An:2012eh}
{\bf DAYA-BAY Collaboration} Collaboration, F.~An {\em et.~al.}, {\it
  {Observation of electron-antineutrino disappearance at Daya Bay}},
  \href{http://arXiv.org/abs/1203.1669}{{\tt 1203.1669}}.

\bibitem{Ahn:2012nd}
{\bf RENO collaboration} Collaboration, J.~Ahn {\em et.~al.}, {\it {Observation
  of Reactor Electron Antineutrino Disappearance in the RENO Experiment}},
  \href{http://arXiv.org/abs/1204.0626}{{\tt 1204.0626}}.

\bibitem{Notzold:87}
D.~Notzold and G.~Raffelt, {\it {Neutrino Dispersion at Finite Temperature and
  Density}},  {\em Nucl. Phys.} {\bf B307} (1988) 924.

\end{thebibliography}
\end{document}